\documentstyle[12pt,epsfig]{article}
\makeatletter
\def\slash#1{{\mathpalette\c@ncel{#1}}} 
\makeatother

\newcommand\beq{\begin{eqnarray}}
\newcommand\eeq{\end{eqnarray}}

\newcommand\la{\langle}
\newcommand\ra{\rangle}

\def\pslash{\rlap/{\mkern-1mu p}}

\def\xhat{\widehat{x}}
\def\zhat{\widehat{z}}

\begin{document}
\begin{flushright}
\end{flushright}
\vspace*{15mm}
\begin{center}
{\Large \bf Contribution of Twist-3 Multi-Gluon
Correlation Functions to Single Spin Asymmetry 
\\[2mm] 
in Semi-Inclusive Deep Inelastic Scattering}
\vspace{1.5cm}\\
 {\sc Hiroo Beppu$^1$, Yuji Koike$^2$, Kazuhiro Tanaka$^3$ and Shinsuke Yoshida$^1$}
\\[0.4cm]
\vspace*{0.1cm}{\it $^1$ Graduate School of Science and Technology, Niigata University,
Ikarashi, Niigata 950-2181, Japan}\\
\vspace*{0.1cm}{\it $^2$ Department of Physics, Niigata University,
Ikarashi, Niigata 950-2181, Japan}\\
\vspace*{0.1cm}{\it $^3$ Department of Physics, 
Juntendo University, Inzai, Chiba 270-1695, Japan}
\\[3cm]

{\large \bf Abstract} \end{center}
As a possible source of the single transverse spin asymmetry,
we study the contribution from purely gluonic correlation
represented by the twist-3 ``three-gluon correlation"
functions in the transversely polarized nucleon.  
We first define a complete set of the relevant three-gluon correlation functions,
and then derive its contribution to the twist-3 single-spin-dependent
cross section for the $D$-meson production in semi-inclusive deep inelastic scattering,
which is relevant to determine the three-gluon correlations.  
Our cross-section formula 
differs from the corresponding result in the literature, and
the origin of the discrepancy is clarified.

\newpage
\section{Introduction}
Clarifying the origin of the large single spin asymmetries (SSAs) observed in 
various high-energy semi-inclusive processes\,\cite{Lambda}-\cite{BRAHMS} has been a big challenge 
during the past decades.  (See \cite{Review} for a review.)
They are now understood as direct consequences of the orbital motion of quarks and gluons and/or
the multi-parton correlations inside the hadrons, and thus 
provide a new opportunity for revealing the QCD dynamics and 
hadron structures,
which do not appear in the parton models and perturbative QCD at the conventional twist-2 level. 
Satisfactory formulation of these effects in QCD, however, 
requires sophisticated framework for which lots of technical developments are involved.  
Up to now, the SSAs have been formulated based on the (naively) ``T-odd" distribution/fragmentation
functions\,\cite{Sivers}-\cite{TMDsidis} within the transverse-momentum-dependent (TMD) factorization
approach\,\cite{CS81,CSS85,JMY05}, and on the twist-3 multi-parton correlation functions in the
collinear factorization approach\,\cite{ET82}-\cite{Kanazawa10}. 
These two mechanisms, in principle, cover different kinematic regions,
such that
the former approach is suitable for
treating SSAs in the region of the small transverse momentum of a particle observed in the final-state, while the latter
is designed for systematic description of SSAs 
at large values of the corresponding transverse momentum.
On the other hand, in the intermediate region of the corresponding transverse momentum 
where both approaches are 
valid,
it's been shown for some structure functions in semi-inclusive deep inelastic scattering (SIDIS) 
and in the Drell-Yan
cross section that 
the two approaches
provide an equivalent description of SSAs\,\cite{JQVY06,JQVY06DIS,KVY08,YZ09}.
Therefore, in practice, the two approaches, tied together at the intermediate transverse-momentum region,
can be regarded as providing
a unique QCD effect leading to SSAs
over the entire kinematic regions.  
Although our understanding on SSAs has made a great progress armed with these mechanisms,  
further studies are yet to be done for a complete clarification
of all the effects responsible for SSAs.  
Among such effects, in particular, the role of purely gluonic effects has not been widely studied in the 
literature.  Since the gluons are ample in the nucleon, they are potentially an
important source of SSAs.  

In this paper, we study a purely gluonic effect as an origin of SSAs in the framework of
the collinear factorization.  
To this end, we work on SSAs for a heavy meson  
production
in SIDIS, in particular, the $D$-meson 
production, $ep\to eDX$.  
Since 
this process is induced dominantly by the 
$c\bar{c}$-pair creation through photon-gluon fusion, this is the most relevant process
to probe the gluonic effects for SSAs, together with $D$-meson production
in $pp$ collisions\,\cite{Anselmino:2004nk,KQVY08} ongoing at RHIC\,\cite{Liu:2009zzw}.  
In the collinear factorization framework, SSA is a twist-3 observable and thus 
the gluonic effects responsible for SSAs are represented by the twist-3 gluon correlation functions
in the polarized nucleon, which were first introduced in the most general form by Ji\,\cite{Ji92}.
Recently the authors of \cite{KQ08} studied the SSAs in SIDIS, $ep^\uparrow\to eDX$,
applying the twist-3 mechanism.  They derived a formula,
in the leading order with respect to the QCD coupling constant, for the
contribution to 
the single-spin-dependent cross section from a ``three-gluon" correlation function.
In our opinion, however, the starting formula used to derive the cross section
in \cite{KQ08} is incomplete 
and cannot
lead to the correct result.
So we revisit the same issue in this paper.  

As is well-known, a twist-2 gluon distribution inside the nucleon is defined in terms of
the gauge-invariant lightcone correlation function
of the gluon field-strength tensors, schematically written as
$\la F_a^{\alpha+}F_{a}^{\beta+}\ra$.
Likewise, the twist-3 ``three-gluon distribution'' functions are defined through
the gauge-invariant correlation functions 
$\la C_\pm^{abc}F_a^{\alpha+}F_{b}^{\beta+}F_{c}^{\gamma+}\ra$,
with the structure constants of the color SU(3) group, $C_+^{abc}=if^{abc}$ and $C_-^{abc}=d^{abc}$.  
On the other hand, when we derive
the three-gluon contribution to the
cross section for $ep^\uparrow\to eDX$ with the large transverse momentum of the $D$ meson, 
we start our analysis 
with a cut forward-amplitude for the cross section,
in which the correlation functions of the gluon fields
appear in the form $\sim\la A_a^{\alpha}A_{b}^{\beta}A_{c}^{\gamma}\ra$. 
Extraction of the twist-3 effect relevant for SSAs from the corresponding amplitude,
converting eventually the associated nucleon matrix elements from 
$\la A_a^{\alpha}A_{b}^{\beta}A_{c}^{\gamma}\ra$
into the gauge-invariant forms $\la C_\pm^{abc} F_a^{\alpha+}F_{b}^{\beta+}F_{c}^{\gamma+}\ra$, is a highly nontrivial issue,
unlike straightforward calculations of the twist-2 cross sections.  
In this connection, it is worth mentioning that,
for the case where the twist-3 quark-gluon correlation functions participate,
the necessary formulation was achieved in \cite{EKT07}; there,
Ward identities played a crucial role to prove the factorization property and
gauge invariance of the corresponding twist-3 cross section. 
Similarly, we will show that, owing to the constraints from the tree-level Ward identities
satisfied by the relevant partonic hard parts, the twist-3 contribution
to the cross section, associated with 
the three-gluon correlation functions $\la A_a^{\alpha}A_{b}^{\beta}A_{c}^{\gamma}\ra$,
can be recast into
the factorized expression 
in terms of the gauge-invariant functions
$\la C_\pm^{abc}F_a^{\alpha+}F_{b}^{\beta+}F_{c}^{\gamma+}\ra$.  
With this formalism we will derive the complete
single-spin-dependent cross section arising from the
twist-3 three-gluon correlation functions of the nucleon.  
We will also clarify the difference of our result from that of \cite{KQ08}.

The remainder of this paper is organized as follows:
In section 2, we first define a complete set of the three-gluon correlation functions
in the transversely polarized nucleon.  
We show that,
actually, the three-gluon correlation functions defined in \cite{Ji92}
are not all independent, i.e., some of them
are redundant.
So we newly define a genuine complete set
of the three-gluon correlation functions.  
In section 3, we present our formalism for calculating
the twist-3 single-spin-dependent cross section for $ep^\uparrow\to eDX$.
We show that only a pole contribution of an 
internal propagator
in the hard part leads to a real quantity relevant to the cross section, 
and that the tree-level
Ward identities satisfied by the corresponding pole contributions play an essential role
to give the factorized expression for the single-spin-dependent cross section, in terms of 
a complete set of the
gauge-invariant correlation functions defined in section 2.
In section 4, we present the final form  of the single-spin-dependent cross section
for $ep^\uparrow\to eDX$
and discuss its characteristic features. 
Section 5 is devoted to a brief summary of our result.
In Appendix A, we summarize the relevant symmetry 
properties of the
gluon correlation functions, which are used in section~3.

\section{Three-gluon correlation functions in the transversely polarized nucleon}
As a straightforward extension of the 
quark-gluon
correlation functions discussed in, e.g., \cite{EKT06,EKT07}, 
purely gluonic correlation
functions can be defined as nucleon matrix elements
of the three gluon fields on the lightcone\,\cite{Ji92}. 
Due to the different ways for contracting the color indices of the three gluon fields 
to obtain the color-singlet operators, one can define the 
two-types of correlation functions as
\beq
&&\hspace{-0.8cm}O^{\alpha\beta\gamma}(x_1,x_2)
=-gi^3\int{d\lambda\over 2\pi}\int{d\mu\over 2\pi}e^{i\lambda x_1}
e^{i\mu(x_2-x_1)}\la pS|d^{bca}F_b^{\beta n}(0)F_c^{\gamma n}(\mu n)F_a^{\alpha n}(\lambda n)
|pS\ra, 
\label{gluond}\\
&&\hspace{-0.8cm}N^{\alpha\beta\gamma}(x_1,x_2)
=-gi^3\int{d\lambda\over 2\pi}\int{d\mu\over 2\pi}e^{i\lambda x_1}
e^{i\mu(x_2-x_1)}\la pS|if^{bca}F_b^{\beta n}(0)F_c^{\gamma n}(\mu n)F_a^{\alpha n}(\lambda n)
|pS\ra, 
\label{gluonf}
\eeq
where $F_a^{\alpha n}\equiv F_a^{\alpha \beta}n_{\beta}$ 
with $F_a^{\alpha\beta}=\partial^\alpha A^\beta_a
-\partial^\beta A^\alpha_a +gf_{abc}A_b^\alpha A_c^\beta$ being the gluon field strength tensor,
$d^{bca}$ and $f^{bca}$ are, respectively, the symmetric
and anti-symmetric structure constants of the color SU(3) group,
and we have suppressed the gauge-link operators which appropriately connect the field strength 
tensors so as to ensure
the gauge invariance.
$p$ is the nucleon momentum, and
$S$ is the transverse spin vector of the
nucleon normalized as $S^2=-1$.
We obtain the twist-3 contributions of (\ref{gluond}) and (\ref{gluonf}), 
when we regard all the free Lorentz indices $\alpha$, $\beta$,
and $\gamma$ to be transverse,
and, in this twist-3 accuracy,
$p$ can be regarded as lightlike ($p^2=0$).
$n$ is another lightlike vector satisfying $p\cdot n=1$, and,   
to be specific, we assume $p^\mu=(p^+,0,\mathbf{0}_\perp)$ and $n^\mu=(0,n^-, \mathbf{0}_\perp)$;
then, we have $S^\mu =(0,0, \mathbf{S}_\perp)$.

Taking into account the constraints from 
hermiticity, 
invariance 
under the parity and time-reversal transformations,
and the permutation symmetry among the participating three gluon-fields,
Ji decomposed the twist-3 contribution of (\ref{gluond}) in terms of the
real, Lorentz-scalar functions $O_1$ and $\widetilde{O}_1$ associated with
the six tensor structures as\,\cite{Ji92}
\beq
&&\qquad
2iM_N\left[
O_1(x_1,x_2)g^{\alpha\beta}\epsilon^{\gamma pnS}
+O_1(x_2,x_2-x_1)g^{\beta\gamma}\epsilon^{\alpha pnS}
+O_1(x_1,x_1-x_2)g^{\gamma\alpha}\epsilon^{\beta pnS}\right.\nonumber\\
&&\qquad\left.
+\widetilde{O}_1(x_1,x_2)\epsilon^{\alpha\beta pn}S^\gamma
+\widetilde{O}_1(x_2,x_2-x_1)\epsilon^{\beta\gamma pn}S^\alpha
-\widetilde{O}_1(x_1,x_1-x_2)\epsilon^{\gamma\alpha pn}S^\beta
\right], \label{JiO}
\eeq
where we have introduced the nucleon mass $M_N$ in order to define 
$O_1$ and $\widetilde{O}_1$ as dimensionless,
and the similar decomposition of (\ref{gluonf}) was also introduced as\,\cite{Ji92}
\beq
&&\qquad
2iM_N\left[
N_1(x_1,x_2)g^{\alpha\beta}\epsilon^{\gamma pnS}
-N_1(x_2,x_2-x_1)g^{\beta\gamma}\epsilon^{\alpha pnS}
-N_1(x_1,x_1-x_2)g^{\gamma\alpha}\epsilon^{\beta pnS}\right.\nonumber\\
&&\qquad\left.
+\widetilde{N}_1(x_1,x_2)\epsilon^{\alpha\beta pn}S^\gamma
-\widetilde{N}_1(x_2,x_2-x_1)\epsilon^{\beta\gamma pn}S^\alpha
+\widetilde{N}_1(x_1,x_1-x_2)\epsilon^{\gamma\alpha pn}S^\beta
\right],
\label{JiN}
\eeq
with the other dimensionless, real functions 
$N_1$ and $\widetilde{N}_1$.
In \cite{Ji92}, the above four functions
$O_1$, $\widetilde{O}_1$, $N_1$ and $\widetilde{N}_1$
were treated as independent twist-3 three-gluon correlation functions.
However, the six tensor structures in (\ref{JiO}) and (\ref{JiN}) are not all
independent and thus  
these decompositions actually
define redundant correlation functions.\footnote{This point was also noticed 
and briefly mentioned in \cite{BJLO}.}
To see this, we recall the identity,
\beq
g^{\mu\nu}\epsilon^{\alpha\beta\rho\delta}
=g^{\mu\alpha}\epsilon^{\nu\beta\rho\delta}
+g^{\mu\beta}\epsilon^{\alpha\nu\rho\delta}
+g^{\mu\rho}\epsilon^{\alpha\beta\nu\delta}
+g^{\mu\delta}\epsilon^{\alpha\beta\rho\nu},
\eeq
and contract both sides 
of this identity 
with the tensor
$g_{\mu}^{\gamma} S_\nu p_\rho n_\delta$.
When the indices $\alpha$, $\beta$ and $\gamma$ are
associated with the transverse components,
one obtains the relation,
\beq
\epsilon^{\alpha\beta pn}S^\gamma=
-g^{\gamma\alpha}\epsilon^{\beta pnS} + g^{\beta\gamma}\epsilon^{\alpha pnS},
\eeq
which shows that the 
tensor structures associated with $\widetilde{O}_1$ and $\widetilde{N}_1$ 
are reexpressed by those associated with 
${O}_1$ and $N_1$, respectively, in (\ref{JiO}) and (\ref{JiN}).  
Accordingly, 
(\ref{gluond}) and (\ref{gluonf})
should be decomposed into the three tensor structures.
This can be formally achieved by omitting the terms associated with 
$\widetilde{O}_1$ and $\widetilde{N}_1$ in (\ref{JiO}) and (\ref{JiN}),
respectively,
and 
we newly define the two twist-3 functions 
$O(x_1,x_2)$ and $N(x_1,x_2)$
as independent three-gluon correlation functions
to represent (\ref{gluond}) and (\ref{gluonf}) as
\beq
&&\hspace{-0.5cm}O^{\alpha\beta\gamma}(x_1,x_2)\nonumber\\
&&=2iM_N\left[
O(x_1,x_2)g^{\alpha\beta}\epsilon^{\gamma pnS}
+O(x_2,x_2-x_1)g^{\beta\gamma}\epsilon^{\alpha pnS}
+O(x_1,x_1-x_2)g^{\gamma\alpha}\epsilon^{\beta pnS}\right]
\label{3gluonO},\\
&&\hspace{-0.5cm}N^{\alpha\beta\gamma}(x_1,x_2)\nonumber\\
&&=2iM_N\left[
N(x_1,x_2)g^{\alpha\beta}\epsilon^{\gamma pnS}
-N(x_2,x_2-x_1)g^{\beta\gamma}\epsilon^{\alpha pnS}
-N(x_1,x_1-x_2)g^{\gamma\alpha}\epsilon^{\beta pnS}\right].
\label{3gluonN}
\eeq
In this paper, we use these $O(x_1,x_2)$ and $N(x_1,x_2)$ 
as constituting a genuine complete set 
to express the 
contribution of the three-gluon correlations to the twist-3 single-spin-dependent
cross section for $ep^\uparrow\to eDX$. 
Similarly to the decomposition (\ref{JiO}) and (\ref{JiN}) considered in \cite{Ji92},
the independent tensor structures in (\ref{3gluonO}) and (\ref{3gluonN})
are parameterized by the common real functions $O$ and $N$, respectively,
as consequences of the constraints from hermiticity, 
invariance 
under the parity and time-reversal transformations,
and the permutation symmetry among the participating three gluon-fields.
In particular,
these constraints imply
the following symmetry relations for $O(x_1,x_2)$ and $N(x_1,x_2)$,
which are associated with the $C$-odd and $C$-even combinations of the three gluon operators
in (\ref{gluond}) and (\ref{gluonf}), respectively:
\beq
&&O(x_1,x_2)=O(x_2,x_1),\qquad O(x_1,x_2)=O(-x_1,-x_2),\label{symO}\\
&&N(x_1,x_2)=N(x_2,x_1),\qquad N(x_1,x_2)=-N(-x_1,-x_2).\label{symN}  
\eeq
The authors of \cite{Braun09} also pointed out that there
are only two independent three-gluon correlation functions at twist-3,
in the study of the 
evolution equations for the twist-3 distributions in the transversely polarized nucleon.  
We also mention that the $C$-even three-gluon correlation function in (\ref{3gluonN})
contributes to the twist-3 transverse-spin 
structure function $g_2(x, Q^2)$ in the inclusive DIS with polarized beam and target~\cite{g2},
and the twist-3 local operators associated with the double moments of (\ref{gluonf}) 
have been analyzed in the framework of the operator
product expansion~\cite{ope}.

As we will see in the next section,
a contribution to the single-spin-dependent cross section 
based on the corresponding factorization formula at the leading order in QCD perturbation theory
is generated from a pole arising in
the propagators
in the hard partonic subprocesses, and such a pole contribution fixes the momentum fractions
in the correlation functions
$O^{\alpha\beta\gamma}(x_1,x_2)$ and
$N^{\alpha\beta\gamma}(x_1,x_2)$
at $x_1=x_2 \equiv x$.  
This represents the situation where
one of the three gluon-lines participating in the partonic subprocesses has zero 
momentum (see (\ref{gluond}), (\ref{gluonf})),
and thus the corresponding contributions relevant to SSA can be referred 
to as the soft-gluon-pole (SGP) contributions. 
Combined with the above decompositions (\ref{3gluonO}) and (\ref{3gluonN}),
the single-spin-dependent cross section
proves to
receive contributions associated with $O(x,x)$, $O(x,0)$, $N(x,x)$ and $N(x,0)$.
In particular, we will 
observe in the next section that the partonic hard part
associated with $O(x,x)$ is different from the hard part 
associated with $O(x,0)$,
due to the difference in the tensor structures among 
the three terms in the right-hand side of (\ref{3gluonO}).
Similarly, the partonic hard part for 
$N(x,x)$ is different from that for
$N(x,0)$, reflecting the difference in the corresponding tensor structures in~(\ref{3gluonN}). 
 
Here, we comment on the treatment of the three-gluon correlation functions 
adopted in the recent work\,\cite{KQ08} on the same phenomenon
as discussed in the present paper, i.e., 
on the twist-3 mechanism to SSAs in SIDIS,
$ep^\uparrow\to eDX$. 
The three-gluon correlation functions introduced by the authors of \cite{KQ08} 
read
\beq
T_G^{(\pm)}(x,x)=\int{dy_1^-dy_2^-\over 2\pi}
e^{ixp^+y_1^-} {1\over xp^+} g_{\beta\alpha}\epsilon_{S\gamma n p}\la pS| 
C^{bca}_\pm F_{b}^{\beta +}(0)F_{c}^{\gamma +}(y_2^-)F_a^{\alpha +}(y_1^-)
|pS\ra, 
\label{3gluonKQ}
\eeq
with 
\begin{equation}
C_+^{bca}=if^{bca}, \;\;\;\;\;\;\;\;\;\;\;\;\;\;\;\;\;\;\;\; C_-^{bca}= d^{bca},
\label{cpm}
\end{equation}
and it was claimed that the corresponding twist-3 single-spin-dependent cross section 
at the leading order in QCD perturbation theory can be expressed entirely
in terms of these two functions of $x$ (see (\ref{KQ}) below and the discussion following 
this formula). 
However, 
$T^{(\pm)}_G(x,x)$ in 
(\ref{3gluonKQ}) can be obtained by the contraction of
$O^{\alpha\beta\gamma}(x,x)$ and
$N^{\alpha\beta\gamma}(x,x)$ in (\ref{gluond}) and (\ref{gluonf})
with the particular tensor $g_{\beta\alpha}\epsilon_{S\gamma n p}$, 
and thus can be written,
using (\ref{3gluonO}) and (\ref{3gluonN}), as
\beq
&&{xg\over 2\pi} T_G^{(+)}(x,x) = -4M_N\left(N(x,x) -N(x,0) \right),\\
&&{xg\over 2\pi} T_G^{(-)}(x,x) = -4M_N\left(O(x,x) +O(x,0) \right). 
\eeq
Using these relations, the 
twist-3 single-spin-dependent cross section obtained in \cite{KQ08} 
implies the same 
partonic hard parts for $O(x,x)$ and $O(x,0)$, and similarly for $N(x,x)$ and $N(x,0)$.
Such result implied by \cite{KQ08} disagrees with our result as
mentioned above: 
In the twist-3 single-spin-dependent cross section induced by the three-gluon correlations,
$O(x,x)$ and $O(x,0)$ are associated with the different partonic hard parts,
and similarly for $N(x,x)$ and $N(x,0)$,
so that the corresponding cross section cannot be expressed by
$T_G^{(\pm)}(x,x)$ only.
In this connection, we also note that
(\ref{3gluonO}) and (\ref{3gluonN})
can be converted to give $O(x,x)$, $O(x,0)$, $N(x,x)$ and $N(x,0)$ 
as
\beq
&&O(x,x)={i\over 8M_N}\left( 3g_{\alpha\beta}\epsilon_{\gamma pnS}
-2g_{\alpha\gamma}\epsilon_{\beta pnS}\right) O^{\alpha\beta\gamma}(x,x),
\label{oxx}\\
&&O(x,0)={-i\over 8M_N}\left( g_{\alpha\beta}\epsilon_{\gamma pnS}
-2g_{\alpha\gamma}\epsilon_{\beta pnS}\right) O^{\alpha\beta\gamma}(x,x),\\
&&N(x,x)={i\over 8M_N}\left( 3g_{\alpha\beta}\epsilon_{\gamma pnS}
-2g_{\alpha\gamma}\epsilon_{\beta pnS}\right) N^{\alpha\beta\gamma}(x,x),\\
&&N(x,0)={i\over 8M_N}\left( g_{\alpha\beta}\epsilon_{\gamma pnS}
-2 g_{\alpha\gamma}\epsilon_{\beta pnS}\right) N^{\alpha\beta\gamma}(x,x)
\label{nx0}.
\eeq
From these relations, we see that the contributions of certain types of 
the twist-3 components 
in the three-gluon correlation functions,
obtained as the contractions of 
$O^{\alpha\beta\gamma}(x,x)$,
$N^{\alpha\beta\gamma}(x,x)$ 
with the tensor $g_{\alpha\gamma}\epsilon_{\beta pnS}$,
were not taken into account in \cite{KQ08}.
Taking into account the components contracted with $g_{\alpha\gamma}\epsilon_{\beta pnS}$,
as well as those contracted with $g_{\alpha\beta}\epsilon_{\gamma pnS}$,
is required on general grounds by the Bose statistics of the gluon,
and, in (\ref{oxx})-(\ref{nx0}),
there is no reason to anticipate that the latter types of components
are more
important than the former.

\section{Formalism for the twist-3 mechanism from three-gluon correlations}

\subsection{Kinematics for $ep^\uparrow\to eDX$}
Here we summarize the kinematics for the SIDIS process,
\beq
e(\ell)+p^\uparrow(p, S) \to e(\ell')+D(P_h)+X. 
\label{SIDIS}
\eeq
As noted in the last section, 
one can assume that the initial nucleon's momentum $p$ is lightlike, $p^2=0$,
in the twist-3 accuracy.  
But, we keep the mass $m_h$ 
of the final charmed hadron ($D$-meson) as 
$P_h^2=m_h^2$. 
The corresponding results for the case of the
light-meson production can be obtained by the replacement
$m_h \rightarrow 0$
in all the formulae below.  
For the process~(\ref{SIDIS}),  
besides the masses of the participating particles,
there are five independent Lorentz invariants as
\beq
S_{ep}=(p+\ell)^2,\quad
x_{bj}={Q^2\over 2p\cdot q},\quad Q^2=-q^2=-(\ell-\ell')^2,
\quad z_f={p\cdot P_h\over p\cdot q},\quad
q_T=\sqrt{-q_t^2}.  
\eeq
Here, $q_t$ is the ``transverse" component of $q$ 
defined as
\beq
q_t^\mu= q^\mu+\left(\frac{m_h^2p\cdot q}{\left(p\cdot P_h\right)^2} 
- {P_h\cdot q\over p\cdot P_h}\right)p^\mu -{p\cdot q\over p\cdot P_h} P_h^\mu,
\eeq
satisfying 
$q_t\cdot p=q_t\cdot P_h=0$.  
In the actual calculation we work in
the ``hadron frame''\,\cite{MOS92} where the virtual photon and the initial nucleon are
collinear, i.e., both move along the $z$-axis. 
In this frame, specifically, their momenta $q$ and $p$ are
given as
\beq
q^\mu = (q^0, \vec{q})=(0,0,0,-Q) \; ,
\eeq
and, similarly,
\beq
p^\mu = \left( {Q\over 2x_{bj}},0,0,{Q\over 2x_{bj}}\right) \; ,
\label{pmu}
\eeq
and the outgoing $D$-meson is assumed to reside in the $xz$ plane:
\beq
P_h^\mu = {z_f Q \over 2}\left( 1 + {q_T^2\over Q^2}+ {m_h^2\over z_f^2Q^2},{2 q_T\over Q},0,
-1+{q_T^2\over Q^2}+{m_h^2\over z_f^2Q^2}\right) \; . 
\label{Dmomem}
\eeq
The transverse momentum of the $D$-meson in this 
frame is given by $P_{hT}=z_f q_T$, which is true in any frame where
the
3-momenta $\vec{q}$ and $\vec{p}$ 
are collinear.
The lepton momentum in this frame can be parameterized as
\beq
\ell^\mu={Q\over 2}\left( \cosh\psi,\sinh\psi\cos\phi,
\sinh\psi\sin\phi,-1\right) \; ,\nonumber\\
\ell'^\mu={Q\over 2}\left( \cosh\psi,\sinh\psi\cos\phi,
\sinh\psi\sin\phi,1\right) \; ,
\label{eq2.lepton}
\eeq
where
\beq
\cosh\psi = {2x_{bj}S_{ep}\over Q^2} -1 \; .
\label{eq2.cosh}
\eeq
We parameterize the transverse spin vector of the initial nucleon $S^\mu$
as
\beq
S^\mu= (0,\cos\Phi_S,\sin\Phi_S,0),
\label{phis}
\eeq
where $\Phi_S$ represents the azimuthal angle of $\vec{S}$ measured from
the hadron plane.
With the above definition, 
the cross section for $ep^\uparrow\to eD X$ can be expressed in terms of
$S_{ep}$, $x_{bj}$, $Q^2$, $z_f$, $q_T^2$, $\phi$ and $\Phi_S$ in the hadron
frame.  Note that $\phi$ and $\Phi_S$ are invariant under boosts in the 
$\vec{q}$-direction, so that the cross section presented below is
the same in any frame where $\vec{q}$ and $\vec{p}$ are collinear.

\subsection{Collinear expansion and gauge invariance for the three-gluon contribution}

The differential cross section for $ep^\uparrow\to eDX$ can be obtained as
\beq
d\Delta\sigma={1\over 2S_{ep}}{d^3 \vec{P}_h\over (2\pi)^3 2P_h^0}{d^3\vec{\ell}'\over(2\pi)^3 
2\ell'^0}
{e^4\over q^4}L_{\mu\nu}(\ell,\ell')W^{\mu\nu}(p,q,P_h),
\label{dsigma}
\eeq
where $L_{\mu\nu}(\ell,\ell')=2(\ell_\mu\ell_\nu^{\prime}+\ell_\nu\ell_\mu^{\prime})
-Q^2g_{\mu\nu}$ is the leptonic tensor for the unpolarized electron, and
$W^{\mu\nu}(p,q,P_h)$ is the hadronic tensor.  
In the present study we are interested in the contribution to $W^{\mu\nu}(p,q,P_h)$
from the three-gluon correlation functions for the initial nucleon, 
in which $c$ and $\bar{c}$ are created through the photon-gluon
fusion process and one of them fragments into a $D$ 
($\bar{D}$)
meson.  
The fragmentation function $D(z)$ for a $c$-quark to become 
the $D$-meson with momentum $P_h$
is defined from the corresponding lightcone correlation function as
\beq
\sum_X{1\over 3}\int{d\lambda\over 2\pi}e^{-i\lambda/z}\la 0|\psi_i(0)|D(P_h) X\ra
\la D(P_h) X| \bar{\psi}_j(\lambda w)|0\ra
=\left( \pslash_c +m_c\right)_{ij}D(z)+\cdots,
\label{fragmentationf}
\eeq
where the ellipses denote 
the terms associated with the gamma matrix 
structures which are irrelevant for the present purpose.
In the left-hand side, we have suppressed the gauge-link
operators to be connected to the quark fields, 
as well as the trace over the color indices.
Here, $z$ is the relevant momentum fraction, 
$w$ is the lightlike vector of $O(1/Q)$ satisfying $P_h \cdot w=1$,  
and $p_c$ is the momentum of the $c$ (or $\bar{c}$) quark with mass $m_c$,
such that $p_c^\mu = P_h^\mu / z + rw^\mu$,
with $r= ( m_c^2z - m_h^2/z )/2$ to satisfy $p_c^2=m_c^2$.
At the leading twist-2 accuracy for the quark-fragmentation process,
we set $w^\mu=p^\mu/(P_h\cdot p)$, and,
in the hadron frame, $p_c$ is expressed as
\beq
p_c^\mu = {\zhat Q \over 2}\left( 1 + {q_T^2\over Q^2}+ {m_c^2\over \zhat^2Q^2},{2 q_T\over Q},0,
-1+{q_T^2\over Q^2}+{m_c^2\over \zhat^2Q^2}\right)  ,
\label{charmmom}
\eeq 
where $\zhat={z_f\over z}$. Note that, when we make the replacement, $\sum_X |D(P_h) X\ra
\la D(P_h) X| \rightarrow |p_c'\ra\la p_c'|$ in the left-hand side of (\ref{fragmentationf}), 
with $|p_c'\ra$ the state with a $c$-quark having the momentum $p_c'\equiv \left. p_c\right|_{z \rightarrow z'}$,
the ellipses in the right-hand side vanish 
and $D(z)\rightarrow \delta( 1/z' -1/z)$.
The fragmentation function $D(z)$ of (\ref{fragmentationf})
is factorized from $W_{\mu\nu}$ as
\beq
W_{\mu\nu}(p,q,P_h)=
\int{dz\over z^2}D(z) w_{\mu\nu}\left(p,q,p_c\right) , 
\label{Wmunu}
\eeq
where the summation over the $c$ and $\bar{c}$ quark contributions is implicit.
To extract the twist-3 effect in $w_{\mu\nu}$, 
one needs to analyze the diagrams of the type shown in Fig. 1.  
However, 
as shown in Appendix A, in the leading order with respect to the QCD coupling constant for the partonic hard scattering parts,
the contribution of Fig.~1(a) can not give rise to the single-spin-dependent cross section,
and only Fig.~1(b) contributes to SSA, due to the symmetry properties
of the correlation functions of two and three gluon fields in the polarized nucleon.  
So we shall focus on the analysis of Fig.~1(b) below.  
The goal of our analysis is to show that
all the contributions from Fig. 1(b) in the twist-3 accuracy
can be expressed in terms of the gauge-invariant correlation functions
$O^{\alpha\beta\gamma}(x_1,x_2)$ and 
$N^{\alpha\beta\gamma}(x_1,x_2)$ defined as (\ref{gluond}) and (\ref{gluonf})
in the previous section.  
For this purpose, we work in Feynman gauge and 
apply the collinear expansion to the hard scattering part in Fig.~1(b),  
keeping all the terms contributing in the twist-3 accuracy\,\cite{EKT07}.   
For simplicity in the notation,
we shall henceforth omit the Lorentz indices $\mu$ and $\nu$ 
for the virtual photon in $w_{\mu\nu}\left(p,q,p_c\right)$ of (\ref{Wmunu})
and write it as
$w\left(p,q,p_c\right)$.  
\begin{figure}[t!]
\begin{center}
\epsfig{figure=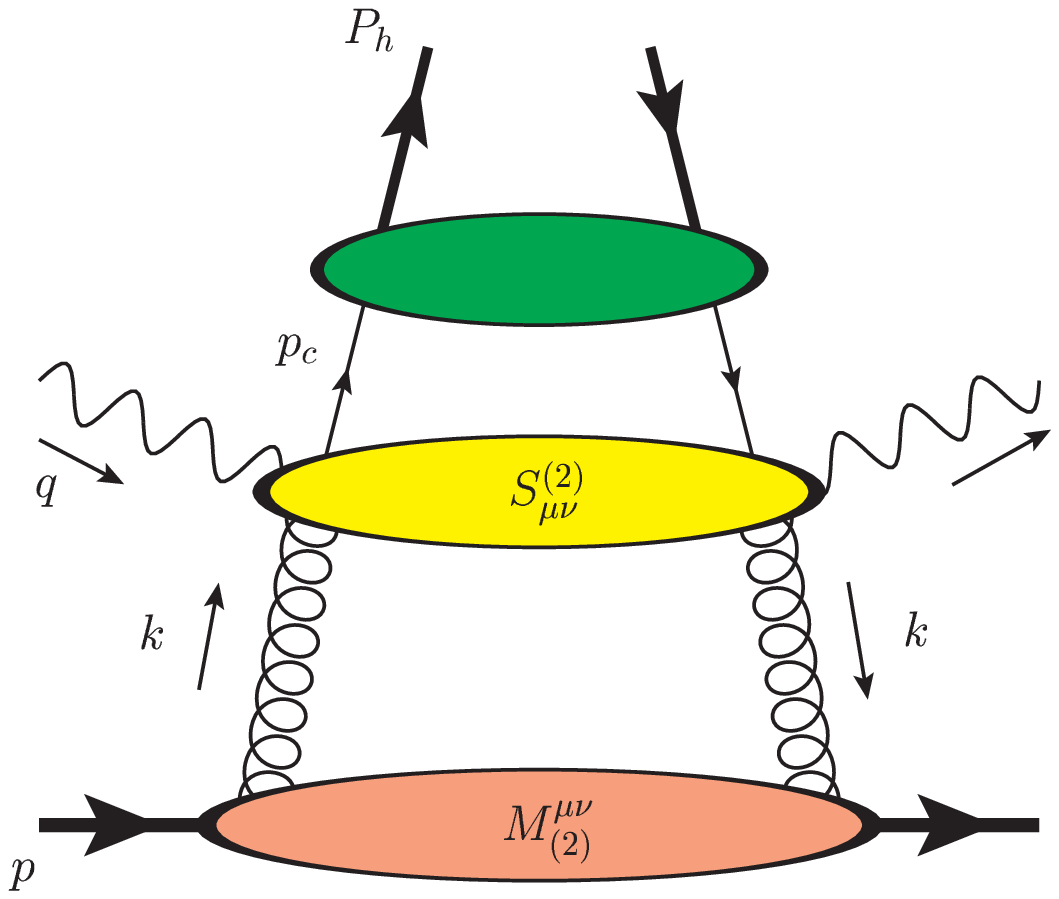,width=0.43\textwidth}
\hspace{1.3cm}
\epsfig{figure=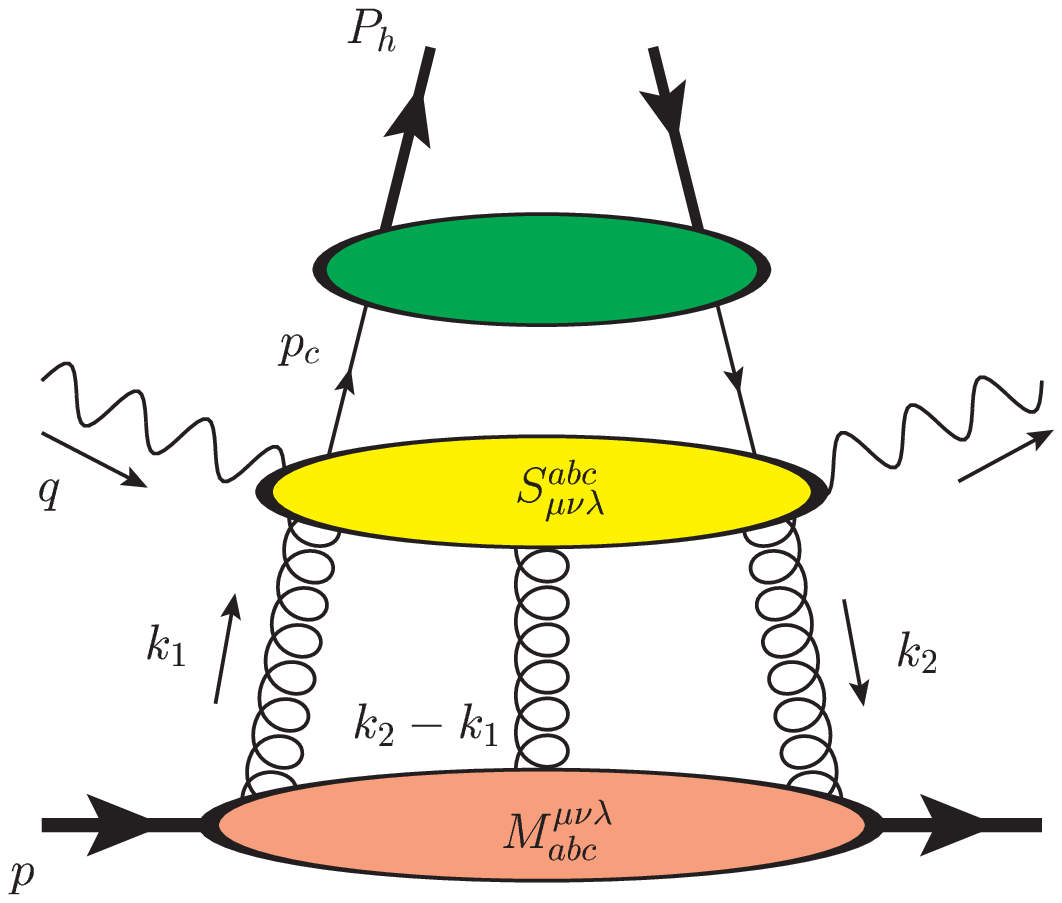,width=0.43\textwidth}
\end{center}
\hspace{3.5cm}(a)\hspace{8cm}(b)
\caption{
Generic diagrams for the hadronic tensor of $ep^\uparrow\to eDX$
induced by the gluonic effect in the nucleon.  Each one is
decomposed into the nucleon matrix element (lower blob), $D$-meson matrix element (upper blob),
and the partonic hard scattering part by the virtual photon (middle blob). 
In the expansion by the number of gluon lines connecting the middle and lower blobs,
the first two terms, (a) and (b),  
are relevant to the twist-3 effect induced by the gluons in the nucleon.
\label{fig1}
}
\end{figure}

The contribution from Fig. 1(b)
to $w(p,q,p_c)$ can be written as
\beq
w\left(p,q,p_c\right)=
\int{d^4k_1\over (2\pi)^4}\int{d^4k_2\over (2\pi)^4}
\,S_{\mu\nu\lambda}^{abc}\left(k_1,k_2,q,p_c\right) 
M^{\mu\nu\lambda}_{abc}(k_1,k_2),
\label{wtensor}
\eeq
where $S_{\mu\nu\lambda}^{abc}(k_1,k_2,q,p_c)$ is the partonic hard scattering
part represented by the middle blob of Fig. 1(b) and $M^{\mu\nu\lambda}_{abc}(k_1,k_2)$
is the corresponding nucleon matrix element (lower blob) defined as
\beq
M^{\mu\nu\lambda}_{abc}(k_1,k_2)=g \int\, d^4\xi\int\, d^4\eta\, e^{ik_1\xi}e^{i(k_2-k_1)\eta}
\la pS | A_b^\nu(0)A_c^\lambda(\eta)A_a^\mu(\xi) |pS\ra.  
\label{Mk1k20}
\eeq
Note that, for later convenience, we include one QCD coupling constant in the definition
of this nucleon matrix element.
In (\ref{wtensor}), a real contribution relevant to the cross section for SSA 
occurs from an imaginary part of the color-projected hard part
$C_\pm^{bca} S^{abc}_{\mu\nu\lambda}(k_1,k_2,q,p_c)$
with (\ref{cpm}),
since $C_\pm^{bca} M^{\mu\nu\lambda}_{abc}(k_1,k_2)$ 
are pure imaginary quantities as shown in Appendix A.  
This means that only the pole contribution produced by an
internal propagator in the hard part 
can give rise to SSA.  

\begin{figure}[t!]
\begin{center}
\epsfig{figure=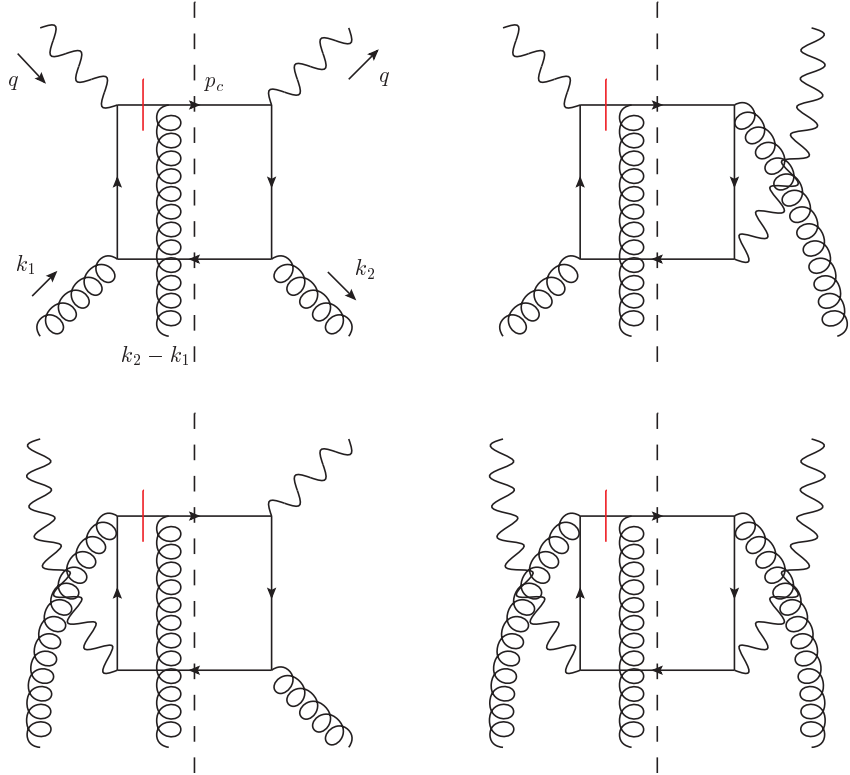,width=0.7\textwidth}
\end{center}
\caption{
Feynman diagrams for the partonic hard part in Fig.~1(b), representing the 
photon-gluon fusion subprocesses that give rise to the ``surviving'' pole contribution for $ep^\uparrow\to eDX$
in the leading order with respect to the QCD coupling constant. 
The short bar on the internal $c$-quark line
indicates that the pole part is to be taken from that propagator.  
In the text, momenta are assigned as shown in the upper-left diagram, where
$p_c$ denotes the momentum of the $c$-quark fragmenting into the $D$-meson
in the final state.  The mirror diagrams also contribute.  
\label{fig2}
}
\end{figure}
In the leading order with respect to the QCD coupling constant, we find that 
four topologically distinct diagrams shown in Fig. 2, together with their
mirror diagrams, give rise to the ``surviving'' pole contributions; 
here, a short bar 
indicates the quark propagator
that produces the corresponding pole contribution.
The other pole contributions
turn out to cancel among themselves
after summing the contributions of all the leading-order diagrams
for (\ref{wtensor}).  
With the assignment of the momenta $k_1$ and $k_2$ of gluons as shown in Fig.~2, 
the condition for those poles is given by $(p_c -k_2 +k_1)^2-m_c^2=0$.
After we perform the collinear expansion 
and reach the collinear limit, $k_i\to x_ip$ ($i=1,\ 2$), with $x_1, x_2$
and $x_2-x_1$
representing the longitudinal momentum fractions of the relevant three gluons,
this condition reduces to $x_1=x_2$
and hence
a pole of such type is referred to as the soft-gluon pole (SGP).  
In the following, we assume that $S_{\mu\nu\lambda}^{abc}\left(k_1,k_2,q,p_c\right)$
in (\ref{wtensor}) represents the sum of the 
contributions of the diagrams in Fig.~2 and their mirror diagrams,
in which 
the barred propagator 
is replaced by its pole contribution.
Also, for simplicity of notation, we suppress the color indices $a,b,c$
and the momenta $q$ and $p_c$ in the hard part 
$S_{\mu\nu\lambda}^{abc}\left(k_1,k_2,q,p_c\right)$, writing it
simply as $S_{\mu\nu\lambda}\left(k_1,k_2\right)$, and
correspondingly, we write $M^{\mu\nu\lambda}_{abc}(k_1,k_2)$ 
as $M^{\mu\nu\lambda}(k_1,k_2)$.

To perform the collinear expansion, we decompose the relevant gluon momenta $k_i$ ($i=1,2$) as
\beq
k_i^\mu=(k_i\cdot n)p^\mu + (k_i\cdot p)n^\mu + k_\perp^\mu
\equiv x_ip^\mu +\omega^\mu_{\ \,\nu}k_i^\nu,
\label{ki}
\eeq
where $x_i= k_i\cdot n$ and $\omega^\mu_{\ \,\nu}\equiv g^\mu_{\ \,\nu}-p^\mu n_\nu$.  
Since $p^\mu \sim g^{\mu}_+ Q$ in the hadron frame with (\ref{pmu}) and thus the component 
along $p^\mu$ 
gives the 
leading contribution in (\ref{ki}) with respect to the hard scale $Q$,
we expand  $S_{\mu\nu\lambda}(k_1,k_2)$ around $k_i=x_ip$.  
Expressing also the gluon field $A^\alpha$ in the Feynman gauge as
\begin{equation}
A^{\alpha}=\left(p^\alpha n_\kappa + \omega^\alpha_{\ \,\kappa}\right) A^\kappa
=p^\alpha n\cdot A + \omega^\alpha_{\ \,\kappa} A^\kappa ,
\label{gluonfield}
\end{equation}
we note that, in the matrix element $M^{\mu\nu\lambda}(k_1,k_2)$ of (\ref{Mk1k20}), 
the
components associated with the second term of (\ref{gluonfield})
give rise to the contributions
suppressed by $\sim 1/Q$ or more, compared with the corresponding contribution due to 
the first term, $p^\alpha n\cdot A$
(see, e.g., \cite{EKT07}).
According to the decomposition (\ref{gluonfield}), 
the integrand of (\ref{wtensor}) can be expressed as
\beq
&&\!\!\!\!
S_{\mu\nu\lambda}\left(k_1,k_2 \right) 
M^{\mu\nu\lambda}(k_1,k_2)
\nonumber\\
&&=S_{\mu\nu\lambda}\left(k_1,k_2 \right) 
(p^\mu n_\kappa+ \omega^\mu_{\ \,\kappa})  (p^\nu n_\tau+ \omega^\nu_{\ \,\tau}) 
(p^\lambda n_\sigma + \omega^\lambda_{\ \,\sigma})   
M^{\kappa \tau \sigma}(k_1,k_2) 
\nonumber\\
&&
=S_{ppp}\left(k_1,k_2 \right) M^{nnn}(k_1,k_2)
+S_{\alpha pp}\left(k_1,k_2 \right) \omega^\alpha_{\ \,\kappa}M^{\kappa nn}(k_1,k_2)
+S_{p\alpha p}\left(k_1,k_2 \right) \omega^\alpha_{\ \,\kappa}M^{n\kappa n}(k_1,k_2)
\nonumber\\
&&+\cdots 
+S_{p \alpha \beta}\left(k_1,k_2 \right) 
\omega^\alpha_{\ \,\kappa}\omega^\beta_{\ \,\tau}M^{n \kappa \tau}(k_1,k_2)
+S_{\alpha \beta \gamma}\left(k_1,k_2 \right) \omega^\alpha_{\ \,\kappa} \omega^\beta_{\ \,\tau}
\omega^\gamma_{\ \,\sigma}M^{\kappa \tau \sigma}(k_1,k_2) ,
\label{wtensorwtensor}
\eeq
where $S_{ppp}(k_1,k_2)\equiv 
S_{\mu\nu\lambda}(k_1,k_2)p^\mu p^\nu p^\lambda$, 
$M^{nnn}(k_1,k_2)\equiv M^{\mu\nu\lambda}(k_1,k_2)n_\mu n_\nu n_\lambda$, etc.
By performing the collinear expansion of the hard part of each term 
in the right-hand side of this formula,
we can organize the integrand of (\ref{wtensor}) 
based on the order counting
with (\ref{ki}), (\ref{gluonfield}),
keeping the terms
necessary in the twist-3 accuracy.
For the first term in the right-hand side of (\ref{wtensorwtensor}), 
the Taylor expansion about $k_i =x_i p$ gives,
\beq
&&\hspace{-0.5cm}S_{ppp}(k_1,k_2)\nonumber\\
&& = S_{ppp}(x_1,x_2)
+\omega^\alpha_{\ \,\kappa}k_1^\kappa  
\left.{\partial S_{ppp}(k_1,k_2) \over \partial k_1^\alpha}\right|_{k_i=x_ip}
+\omega^\alpha_{\ \,\kappa}k_2^\kappa  
\left.{\partial S_{ppp}(k_1,k_2) \over \partial k_2^\alpha}\right|_{k_i=x_ip}\nonumber\\
&& +{1\over 2}\omega^\alpha_{\ \,\kappa}k_1^\kappa \,\omega^\beta_{\ \,\tau}k_1^\tau  
\left.{\partial^2 S_{ppp}(k_1,k_2) \over 
\partial k_1^\alpha\partial k_1^\beta }\right|_{k_i=x_ip}
+{1\over 2}\omega^\alpha_{\ \,\kappa}k_2^\kappa \,\omega^\beta_{\ \,\tau}k_2^\tau  
\left.{\partial^2 S_{ppp}(k_1,k_2) \over 
\partial k_2^\alpha\partial k_2^\beta }\right|_{k_i=x_ip}\nonumber\\
&& +\omega^\alpha_{\ \,\kappa}k_1^\kappa \,\omega^\beta_{\ \,\tau}k_2^\tau  
\left.{\partial^2 S_{ppp}(k_1,k_2) \over 
\partial k_1^\alpha\partial k_2^\beta }\right|_{k_i=x_ip}\nonumber\\
&& +{1\over 6}\omega^\alpha_{\ \,\kappa}k_1^\kappa 
\,\omega^\beta_{\ \,\tau}k_1^\tau \,\omega^\gamma_{\ \,\sigma}k_1^\sigma  
\left.{\partial^3 S_{ppp}(k_1,k_2) \over 
\partial k_1^\alpha \partial k_1^\beta \partial k_1^\gamma }\right|_{k_i=x_ip}
+{1\over 6}\omega^\alpha_{\ \,\kappa}k_2^\kappa 
\,\omega^\beta_{\ \,\tau}k_2^\tau \,\omega^\gamma_{\ \,\sigma}k_2^\sigma  
\left.{\partial^3 S_{ppp}(k_1,k_2) \over 
\partial k_2^\alpha \partial k_2^\beta \partial k_2^\gamma }\right|_{k_i=x_ip}\nonumber\\
&&+{1\over 2}\omega^\alpha_{\ \,\kappa}k_1^\kappa 
\,\omega^\beta_{\ \,\tau}k_1^\tau \,\omega^\gamma_{\ \,\sigma}k_2^\sigma  
\left.{\partial^3 S_{ppp}(k_1,k_2) \over 
\partial k_1^\alpha \partial k_1^\beta \partial k_2^\gamma }\right|_{k_i=x_ip}
+{1\over 2}\omega^\alpha_{\ \,\kappa}k_1^\kappa\, 
\omega^\beta_{\ \,\tau}k_2^\tau \,\omega^\gamma_{\ \,\sigma}k_2^\sigma  
\left.{\partial^3 S_{ppp}(k_1,k_2) \over 
\partial k_1^\alpha \partial k_2^\beta \partial k_2^\gamma }\right|_{k_i=x_ip}\nonumber\\
&&+ \cdots . \label{collinear1}
\eeq
Here and below, we use the notation 
$S_{\mu\nu\lambda}(x_1,x_2)\equiv S_{\mu\nu\lambda}(x_1p,x_2p)$
for simplicity. 
We have written down the expansion explicitly up to the third-order terms.  
This is because,  
according to the above notice with the decompositions (\ref{ki}), (\ref{gluonfield}),
the third-order terms in (\ref{collinear1}) behave as the same order as the first term in 
\beq
\hspace{-0.5cm}
S_{\alpha\beta\gamma}(k_1,k_2)\omega^\alpha_{\ \,\kappa} \omega^\beta_{\ \,\tau}
\omega^\gamma_{\ \,\sigma}M^{\kappa \tau \sigma}(k_1,k_2) 
= \left[ S_{\alpha\beta\gamma}(x_1,x_2)+\cdots\right] \omega^\alpha_{\ \,\kappa} \omega^\beta_{\ \,\tau}
\omega^\gamma_{\ \,\sigma}M^{\kappa \tau \sigma}(k_1,k_2) ,
\label{collinear4}
\eeq
which is obtained by the Taylor expansion of the last term of (\ref{wtensorwtensor}).
If we substitute (\ref{collinear4}) directly into the integrand of (\ref{wtensor})
and perform the integrals over $k_1$ and $k_2$,
the first term in the right-hand side produces the 
contribution, which is associated with 
the hard scattering between the three physical gluons from the nucleon and
behaves as the same order as the formal convolution of $S_{\alpha\beta\gamma}(x_1,x_2)$
with the twist-3 correlation functions of (\ref{gluond}) and (\ref{gluonf}).
Actually, this corresponds to a quantity of twist-3.
Compared to this, the ellipses in (\ref{collinear4}) give rise to the 
terms suppressed by $1/Q$ or more corresponding to twist-4 and higher,  
and thus are irrelevant here.  
We can write down the collinear expansions for the hard parts associated with the 
other terms in (\ref{wtensorwtensor}), similarly as (\ref{collinear1}):
We expand $S_{\alpha pp}(k_1,k_2)$, $S_{p\alpha p}(k_1,k_2)$ and $S_{pp\alpha}(k_1,k_2)$
through the terms involving the second derivatives,
and $S_{\alpha\beta p}(k_1,k_2)$, $S_{\alpha p\beta}(k_1,k_2)$ and $S_{p\alpha\beta}(k_1,k_2)$
through the terms involving the first derivatives.
Thus, the collinear expansion of 
$S_{\mu\nu\lambda}(k_1,k_2)$ in (\ref{wtensor}) produces lots of terms as above, 
and each of those terms is not gauge invariant.
At first sight, it looks hopeless to
reorganize those into a 
form of the convolution
with only the gauge-invariant correlation functions
$O^{\alpha\beta\gamma}(x_1,x_2)$ and 
$N^{\alpha\beta\gamma}(x_1,x_2)$ of (\ref{gluond}) and (\ref{gluonf}) used.

However, as was the case for the pion production 
associated with the twist-3 quark-gluon correlation functions\,\cite{EKT07}, 
great simplification occurs due to
Ward identities satisfied by the 
corresponding partonic hard-scattering function, $S_{\mu\nu\lambda}(k_1,k_2)$.  
To derive the Ward identities, we note that 
the contribution to $S_{\mu\nu\lambda}(k_1,k_2)$ from
each diagram in Fig.~2 all contains the two delta functions,
$\delta\left((k_2+q-p_c)^2-m_c^2\right)$
and $\delta\left( (p_c+k_1-k_2)^2-m_c^2\right)$,
representing the on-shell conditions associated, respectively, with the final-state cut on the unobserved 
$\bar{c}$-quark line and with the unpinched pole contribution of the barred propagator.
Also, in each diagram in Fig.~2, 
the $c$-quark line fragmenting 
into the final-state $D$-meson
is on-shell, see (\ref{charmmom}).
\begin{figure}[t!]
\begin{center}
\epsfig{figure=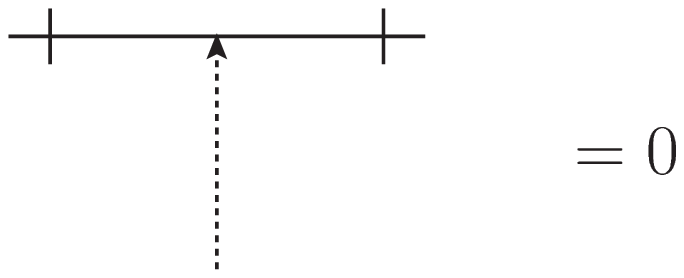,width=0.25\textwidth}
\hspace{3cm}
\epsfig{figure=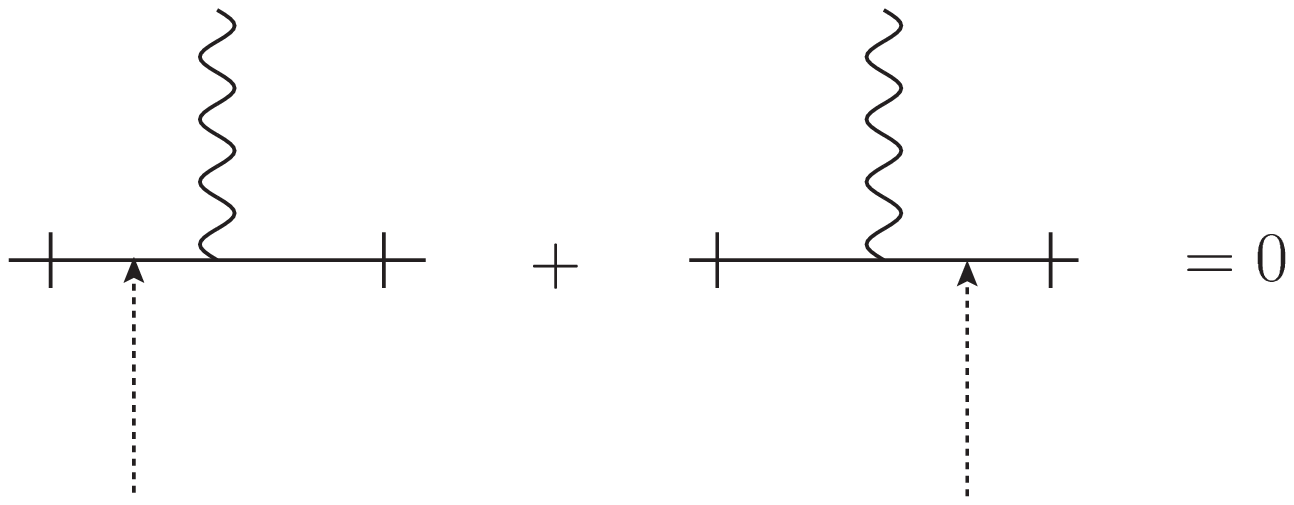,width=0.5\textwidth}
\end{center}
\hspace{2cm}(a)\hspace{9cm}(b)
\caption{Ward identities used for the pole contribution.
The dotted line represents a scalar-polarized gluon, and the quark lines marked by
a bar are on-shell.  
\label{fig3}
}
\end{figure}
Due to these on-shell conditions for the diagrams in Fig.~2
and the similar conditions for their mirror diagrams, 
$S_{\mu\nu\lambda}(k_1,k_2)$ satisfies the tree-level Ward identities,
\beq
&&k_1^\mu S_{\mu\nu\lambda}(k_1,k_2)=0,\nonumber\\
&&k_2^\nu S_{\mu\nu\lambda}(k_1,k_2)=0,\nonumber\\
&&(k_2-k_1)^\lambda S_{\mu\nu\lambda}(k_1,k_2)=0.  
\label{Ward}
\eeq
Here, the last identity and the first two identities are represented diagrammatically in Figs.~3(a) and 3(b), respectively.
In the collinear limit, $k_i\to x_ip$ ($i=1,\ 2$), we have
$\delta\left( (p_c+k_1-k_2)^2-m_c^2\right)\rightarrow (1/2p_c\cdot p) \delta(x_1-x_2)$,
while the other delta function associated with the diagrams in Fig.~2
implies $x_2 >x_{bj}$ (see (\ref{deltadelta}) below);
the similar relations hold also for the corresponding mirror diagrams.
Thus, the collinear expansion of Ward identities of (\ref{Ward}) produces
a series of relations in the collinear limit:
\beq
&&S_{p\nu\lambda}(x_1,x_2)=0,
\label{ward1}\\
&&S_{\mu p\lambda}(x_1,x_2)=0,\label{ward2}\\
&&(x_2-x_1)S_{\mu\nu p}(x_1,x_2)=0, \label{ward3}\\
&&\left. {\partial S_{\mu p \lambda}(k_1,k_2)\over \partial k_1^\alpha}\right|_{k_i=x_ip}=0,\qquad
\left. {\partial S_{p\nu \lambda}(k_1,k_2)\over \partial k_2^\alpha}\right|_{k_i=x_ip}=0,
\label{ward4}\\
&&S_{\alpha\nu \lambda}(x_1,x_2)+
x_1\left.{\partial S_{p\nu \lambda}(k_1,k_2)\over \partial k_1^\alpha}\right|_{k_i=x_ip}
=0,
\label{ward5}
\\
&&S_{\mu\beta \lambda}(x_1,x_2)+
x_2\left.{\partial S_{\mu p \lambda}(k_1,k_2)\over \partial k_2^\beta}\right|_{k_i=x_ip}
=0,
\label{ward6}\\
&&\left.{\partial^2 S_{\mu p\lambda}(k_1,k_2)\over \partial k_1^\alpha
\partial k_1^\beta }\right|_{k_i=x_ip}=
\left.{\partial^2 S_{p\nu \lambda}(k_1,k_2)\over \partial k_2^\alpha
\partial k_2^\beta }\right|_{k_i=x_ip}=0,\label{word67}\\
&&x_2 \left.{\partial^2 S_{\mu p\lambda}(k_1,k_2)\over \partial k_1^\alpha
\partial k_2^\beta }\right|_{k_i=x_ip}
+\left.{\partial S_{\mu\beta \lambda}(k_1,k_2)\over \partial k_1^\alpha}\right|_{k_i=x_ip}=0, 
\label{ward7}
\eeq
and so on, where $\alpha \neq +$, $\beta \neq +$.
These can be used to
reorganize various terms obtained by the collinear expansion of (\ref{wtensorwtensor}),
and one obtains the following results for the contribution in each order:  
\begin{enumerate}
\item[(i)]
By the relations (\ref{ward1}), (\ref{ward2}) and (\ref{ward4}), the first three terms 
in (\ref{collinear1}), and all terms arising in (\ref{wtensorwtensor})
up to the order in $1/Q$ of those three terms, vanish.  
\item[(ii)]
By the relations (\ref{ward3}), (\ref{ward5})-(\ref{ward7}),
the sum of the contributions of the next higher order in  
(\ref{wtensorwtensor}), which behave as the same order as the terms involving the 
second derivative in (\ref{collinear1}), vanish (see the discussion below (\ref{nextstep1})).  
\item[(iii)] 
The remaining contributions in  
(\ref{wtensorwtensor}), behaving as the same order as the first term 
in (\ref{collinear4}), eventually yield the gauge-invariant twist-3 contribution 
to $w(p,q,p_c)$ in (\ref{wfinal}) below.
\end{enumerate}

Instead of describing in detail those rather lengthy calculations
in the framework of the standard collinear expansion,
we may employ a somewhat different approach:
Among the relevant relations (\ref{ward1})-(\ref{ward7}),
in particular,
(\ref{ward5}), (\ref{ward6}) and (\ref{ward7}) play a
role to connect the two terms that are generated 
from the different terms in the right-hand side of (\ref{wtensorwtensor})
through the collinear expansion.
Namely, 
to obtain the gauge-invariant result,
we have to combine the contributions from different terms in (\ref{wtensorwtensor}),
and, furthermore, those terms are associated with different numbers of derivatives.
These facts suggest an approach to apply directly Ward identities of (\ref{Ward}) 
to the second line of (\ref{wtensorwtensor}), before the collinear expansion, 
i.e., without the expansion implied by the second equality in (\ref{wtensorwtensor}), 
nor the Taylor expansion about $k_i =x_i p$.
Indeed, substituting the decomposition (\ref{ki}) into the first two identities in 
(\ref{Ward}),
we obtain
\beq
&&S_{p\nu\lambda}(k_1,k_2)=
-\frac{1}{x_1}\omega^\mu_{\ \,\alpha}k_1^\alpha S_{\mu \nu\lambda}(k_1,k_2),\nonumber\\
&&S_{\mu p\lambda}(k_1,k_2)= -\frac{1}{x_2}\omega^\nu_{\ \,\beta}k_2^\beta S_{\mu \nu\lambda}(k_1,k_2), 
\label{Ward2}
\eeq
and we apply these relations to the second line of (\ref{wtensorwtensor}), yielding
\beq
&& \!\!\!\!\!\!\!
S_{\mu\nu\lambda}\left(k_1,k_2 \right) 
M^{\mu\nu\lambda}(k_1,k_2)
\nonumber\\
&&\!\!\!\!\!\!
=S_{\mu\nu\lambda}\left(k_1,k_2 \right) 
\frac{1}{x_1} \omega^\mu_{\ \,\alpha} \left(-k_1^\alpha n_\kappa+ k_1\cdot n g_\kappa^\alpha\right) 
\nonumber\\
&&\times
\frac{1}{x_2} \omega^\nu_{\ \,\beta} \left(-k_2^\beta n_\tau + k_2\cdot n g_\tau^\beta\right) 
(p^\lambda n_\sigma + \omega^\lambda_{\ \,\sigma})   
M^{\kappa \tau \sigma}(k_1,k_2) ,
\eeq
where some factors combined with the matrix element (\ref{Mk1k20}) 
in the right-hand side can be reexpressed as, 
restoring the color indices $a,b$ and $c$,
\beq
\left(-k_1^\alpha n_\kappa+ k_1\cdot n g_\kappa^\alpha\right)
&&\!\!\!
\left(-k_2^\beta n_\tau + k_2\cdot n g_\tau^\beta\right) 
M^{\kappa \tau \sigma}_{abc}(k_1,k_2) \equiv M^{\alpha \beta \sigma}_{A, abc}(k_1,k_2) 
\nonumber\\
&&=
g \int\, d^4\xi\int\, d^4\eta\, e^{ik_1\xi}e^{i(k_2-k_1)\eta}
\la pS | F_b^{\beta n}(0)A_c^\sigma(\eta)F_a^{\alpha n}(\xi) |pS\ra ,  
\label{Mk1k20M}
\eeq
up to the correction terms beyond the present
lowest-order calculation in QCD perturbation theory, so that
\begin{equation}
S_{\mu\nu\lambda}\left(k_1,k_2 \right) 
M^{\mu\nu\lambda}(k_1,k_2)
=S_{\mu\nu\lambda}\left(k_1,k_2 \right) 
\frac{\omega^\mu_{\ \,\alpha} \omega^\nu_{\ \,\beta}}{x_1 x_2}
\left[p^\lambda M^{\alpha \beta n}_A(k_1,k_2) 
+ \omega^\lambda_{\ \,\sigma}M^{\alpha \beta \sigma}_A(k_1,k_2) \right].
\label{nextstep}
\end{equation}

As the next step, we perform the collinear expansion of the hard-scattering
function.
Based on the definition (\ref{Mk1k20M}) and the decomposition (\ref{gluonfield}),
we note that the first and second terms in the parentheses in (\ref{nextstep})
correspond, respectively, to the ``second'' and ``third'' orders in the order counting
relevant to the collinear expansion like (\ref{collinear1}).
Thus, the collinear expansion of (\ref{nextstep}), up to the desired order,
reads (see the discussion below (\ref{collinear4}))
\beq
&&\!\!\!\!\!\!\!\!\!\!\!\!
S_{\mu\nu\lambda}\left(k_1,k_2 \right) 
M^{\mu\nu\lambda}(k_1,k_2)
=
\frac{\omega^\mu_{\ \,\alpha} \omega^\nu_{\ \,\beta}}{x_1 x_2}\left[
\left(S_{\mu\nu p}\left(x_1,x_2 \right)  +\omega^\lambda_{\ \,\kappa}k_1^\kappa  
\left.{\partial S_{\mu\nu p}(k_1,k_2) \over \partial k_1^\lambda}\right|_{k_i=x_ip}
\right.\right.\nonumber\\
&&\;\;\;\;\;\;\;\;\;
\left.\left.+\omega^\lambda_{\ \,\kappa}k_2^\kappa  
\left.{\partial S_{\mu\nu p}(k_1,k_2) \over \partial k_2^\lambda}\right|_{k_i=x_ip}\right)
M^{\alpha \beta n}_A(k_1,k_2) 
+ S_{\mu\nu \lambda}\left(x_1,x_2 \right)  
\omega^\lambda_{\ \,\sigma}M^{\alpha \beta \sigma}_A(k_1,k_2) \right] .
\label{nextstep1}
\eeq
First, we consider the first term in the right-hand side, which 
is proportional to $S_{\mu\nu p}\left(x_1,x_2 \right)= S_{\mu\nu \lambda}\left(x_1,x_2 \right)p^\lambda$
and corresponds to
the second-order term in the above-mentioned order counting.
One can show that, by the direct diagrammatic calculation of 
$S_{\mu\nu p}\left(x_1,x_2 \right)$,
the corresponding SGP contributions from the diagrams in Fig.~2 
cancel with those from their mirror diagrams 
(i.e., $S_{\mu\nu p}\left(x_1,x_2 \right)$ arising in (\ref{ward3}) equals zero
even for $x_1=x_2$),
and thus 
the first term in the parentheses $\left(\cdots\right)$ in the right-hand side of (\ref{nextstep1})
vanishes.
It is worth noting that the similar vanishing property of the SGP contributions
was used 
also for the case of the pion production 
associated with the twist-3 quark-gluon correlation functions (see, e.g., \cite{EKT07}).
On the other hand, $S_{\mu\nu \lambda}\left(x_1,x_2 \right) \omega^\lambda_{\ \,\sigma}$,
arising in the last term in (\ref{nextstep1}), does not vanish, but this can be recast
using the relation,
\begin{equation}
S_{\mu\nu \lambda}\left(x_1,x_2 \right)\omega^\lambda_{\ \,\sigma}  =
\left.\left(x_1 -x_2\right)\omega^\lambda_{\ \,\sigma}{\partial S_{\mu\nu p}(k_1,k_2)\over \partial k_2^\lambda}\right|_{k_i=x_ip},
\end{equation}
which is obtained by the collinear expansion of the last Ward identity of (\ref{Ward}).
Furthermore, for the second term in the right-hand side of (\ref{nextstep1}),
we may use another relation,
\beq
\left.{\partial S_{\mu\nu p}(k_1,k_2)\over \partial k_1^\lambda }\right|_{k_i=x_ip}=
-\left.{\partial S_{\mu\nu p}(k_1,k_2)\over \partial k_2^\lambda }\right|_{k_i=x_ip},
\eeq
which can be derived for $\lambda=\perp$ by direct inspection of 
the diagrams in Fig.~2 and their mirror diagrams.
We remind that
a similar relation also holds for the hard part corresponding to 
the quark-gluon correlation functions, as discussed 
for the case of the pion production\,\cite{EKT07}.  
Thus, (\ref{nextstep1}) yields, at the twist-3 accuracy,
\beq
S_{\mu\nu\lambda}\left(k_1,k_2 \right) 
M^{\mu\nu\lambda}(k_1,k_2)
&&\!\!\!
=
\frac{\omega^\mu_{\ \,\alpha} \omega^\nu_{\ \,\beta}}{x_1 x_2}
\omega^\lambda_{\ \,\kappa}
\left.{\partial S_{\mu\nu p}(k_1,k_2) \over \partial k_2^\lambda}\right|_{k_i=x_ip}
\nonumber\\
&&\times
\left[
- \left(k_1^\kappa 
-k_2^\kappa \right)
n_\sigma 
+(k_1-k_2)\cdot n
g_\sigma^\kappa
\right] M^{\alpha \beta \sigma}_A(k_1,k_2) ,
\label{nextstep2}
\eeq
where the second line gives, similarly as in (\ref{Mk1k20M}),
\beq
&& \!\!\!\!\!\!\!\!\!\!\!\!
\left[
- \left(k_1^\kappa 
-k_2^\kappa \right)
n_\sigma 
+(k_1-k_2)\cdot n
g_\sigma^\kappa
\right] M^{\alpha \beta \sigma}_{A,abc}(k_1,k_2) 
\nonumber\\
&&=
ig \int\, d^4\xi\int\, d^4\eta\, e^{ik_1\xi}e^{i(k_2-k_1)\eta}
\la pS | F_b^{\beta n}(0)F_c^{\kappa n}(\eta)F_a^{\alpha n}(\xi) |pS\ra ,  
\label{Mk1k20MM}
\eeq
up to the higher-order corrections beyond the present accuracy.
Substituting these results into (\ref{wtensor}),
we obtain the final form for
the relevant twist-3
contribution to
the hadronic tensor in SIDIS,
as the factorization formula
in terms of the gauge-invariant 
three-gluon correlation functions,  
\beq
w\left(p,q,p_c\right)
=\int {dx_1 \over x_1}\int{dx_2\over x_2}
\left. 
{\partial S^{abc}_{\mu\nu\lambda}(k_1,k_2,q,p_c)p^\lambda
\over \partial k_2^\sigma}\right|_{k_i=x_ip}
\omega^\mu_{\ \alpha}\,\omega^\nu_{\ \beta}\,\omega^\sigma_{\ \gamma}\,
{\cal M}^{\alpha\beta\gamma}_{F,abc}(x_1,x_2), 
\label{wfinal}
\eeq
where we have restored the color indices $a,b,c$ as well as momentum variables $q, p_c$,
which were associated with the hard-scattering function in (\ref{wtensor}),
and ${\cal M}^{\alpha\beta\gamma}_{F,abc}(x_1,x_2)$ denote 
the three-gluon lightcone correlation functions,
obtained by integrating (\ref{Mk1k20MM}) over $k_i^-, {\mathbf{k}}_{i\perp}$ ($i=1, 2$),
as
\beq
{\cal M}^{\alpha\beta\gamma}_{F,abc}(x_1,x_2)
&&
=
-gi^3\int{d\lambda\over 2\pi}\int{d\mu\over 2\pi}e^{i\lambda x_1}
e^{i\mu(x_2-x_1)}
\la pS|F_b^{\beta n}(0)F_c^{\gamma n}(\mu n)F_a^{\alpha n}(\lambda n)
|pS\ra
\nonumber\\
&&=\frac{3}{40}d^{abc}O^{\alpha\beta\gamma}(x_1,x_2)
-\frac{i}{24}f^{abc}N^{\alpha\beta\gamma}(x_1,x_2) ,
\label{MFabc}
\eeq
with $O^{\alpha\beta\gamma}(x_1,x_2)$ and 
$N^{\alpha\beta\gamma}(x_1,x_2)$ in (\ref{gluond}) and (\ref{gluonf}).  
As we demonstrated above in deriving these results,
all the gauge-noninvariant terms that could potentially contribute to $w(p,q,p_c)$ 
vanished or canceled among themselves,
and the
{\it total twist-3} contribution to
$w(p,q,p_c)$, relevant to SSA, proves to be expressed solely in terms of the gauge-invariant three-gluon correlation
functions $O(x_1,x_2)$ and 
$N(x_1,x_2)$ defined as (\ref{3gluonO}) and (\ref{3gluonN}).

When one calculates the relevant hard part,
$\left.\partial S^{abc}_{\mu\nu\lambda}(k_1,k_2,q,p_c)p^\lambda
/ \partial k_2^\sigma\right|_{k_i=x_ip}$, 
arising in (\ref{wfinal}),
one should note that
the derivative with respect to $k_2^\sigma$ can hit the delta functions,
$\delta\left((k_2+q-p_c)^2-m_c^2\right)$
and $\delta\left( (p_c+k_1-k_2)^2-m_c^2\right)$, which are 
involved in $S^{abc}_{\mu\nu\lambda}(k_1,k_2,q,p_c)p^\lambda$ as mentioned 
above (\ref{Ward}).
Such derivatives of these delta functions can be 
reexpressed by the derivatives with respect to $x_2$, and then be
treated by integration by parts,
giving rise to the derivative of the three-gluon correlation functions of 
(\ref{3gluonO}) and (\ref{3gluonN}).
After such manipulations, 
the former of the above delta functions,
which represents the  on-shell condition for the final-state cut on the unobserved 
$\bar{c}$-quark line in the diagrams of Fig.~2,
becomes
\beq
\left.\delta\left( (k_2+q-p_c)^2-m_c^2\right)\right|_{k_2=xp} ={1\over \zhat Q^2}
\delta\left( {q_T^2\over Q^2} -\left({1\over \xhat}-1\right)\left({1\over \zhat}-1\right)
+{m_c^2\over \zhat^2 Q^2}\right), 
\label{deltadelta}
\eeq
with $\xhat={x_{bj}\over x}$, and this factor
appears in the final expression for our cross section based on (\ref{wfinal}).

Here, we make a
brief comment on the calculation presented in \cite{KQ08}. 
Using the notation in the present paper,
the authors of \cite{KQ08} calculated $w\left(p,q,p_c\right)$
with the following formula, in place of the right-hand side of (\ref{wfinal}):
\beq
\int {dx_1 \over x_1}\int{dx_2\over x_2}
\left. 
{\partial S^{abc}_{\mu\nu\lambda}(k_1,k_2,q,p_c)p^\lambda g_{\perp}^{\mu\nu}
\over \partial k_{2\perp}^\sigma}\right|_{k_i=x_ip}
\omega^\sigma_{\ \gamma}\,
g_{\alpha\beta}
{\cal M}^{\alpha\beta\gamma}_{F,abc}(x_1,x_2),  
\label{KQ}
\eeq
where 
\beq
g_\perp^{\mu\nu}=g^{\mu\nu}-p^\mu n^\nu -p^\nu n^\mu
=-S^\mu S^\nu - \epsilon^{\mu pnS} \epsilon^{\nu pnS} .
\label{KQ2}
\eeq
In (\ref{KQ}), 
we can make the replacement $\omega^\sigma_{\ \gamma} \rightarrow g_{\perp \gamma}^\sigma$,
up to the irrelevant corrections of twist-4 and higher,
and, using (\ref{KQ2}) and the property
$S_\gamma g_{\alpha\beta} {\cal M}^{\alpha\beta\gamma}_{F,abc}(x_1,x_2)=0$, 
implied by (\ref{MFabc})
with (\ref{3gluonO}) and (\ref{3gluonN}),
we see that the three-gluon correlation functions involved in (\ref{KQ})
are indeed expressed by the two types of functions of (\ref{3gluonKQ}) 
after evaluating the SGP at $x_1=x_2$ explicitly.
Clearly, (\ref{KQ}) used in \cite{KQ08} leads to a result different from 
the result based on our complete formula (\ref{wfinal}).  
It is straightforward to see that, 
if 
the tensor structure of the three-gluon correlation functions 
${\cal M}^{\alpha\beta\gamma}_{F,abc}(x_1,x_2)$ of (\ref{MFabc})
were assumed to be given by only one structure, $g^{\alpha \beta}\epsilon^{\gamma pnS}$,
our formula (\ref{wfinal}) would reduce to the formula (\ref{KQ}),
up to the corrections of twist-4 and higher.
However, such assumption contradicts with the permutation symmetry
required by the Bose statistics of the gluon,
as emphasized in section 2 and represented in (\ref{3gluonO}) and (\ref{3gluonN}).

\subsection{Calculation of $L_{\mu\nu}W^{\mu\nu}$}

Using the kinematic variables defined in section 3.1, 
the differential cross section
(\ref{dsigma}) 
can be expressed
as 
\beq
{d^5\Delta\sigma \over
d x_{bj}dQ^2dz_f dq_T^2 d\phi}
={\alpha_{em}^2 \over 64\pi^3 x_{bj}^2 S_{ep}^2 Q^2}
z_f L^{\mu\nu}(\ell, \ell')W_{\mu\nu}(p,q,P_h),
\label{diffsigma}
\eeq
where 
$\alpha_{em}=e^2/(4\pi)$ is the QED coupling constant.  
We restore the implicit ``free'' Lorentz indices in (\ref{wfinal}) for the virtual photon,
corresponding to $\mu$ and $\nu$ in $w_{\mu \nu}(p,q, p_c)$ of
(\ref{Wmunu}) (see the discussion above (\ref{wtensor})), and, 
substituting the result into (\ref{Wmunu}),
we obtain the three-gluon-correlation contribution 
to $W_{\mu\nu}(p, q, P_h)$ in (\ref{diffsigma}).  
To calculate the contraction $L^{\mu\nu}(\ell, \ell')W_{\mu\nu}(p,q,P_h)$
arising in (\ref{diffsigma}), 
we introduce the following four vectors which are orthogonal to
each other:
\beq
T^\mu &=&{1\over Q}\left(q^\mu + 2x_{bj}p^\mu\right),\nonumber\\
X^\mu &=&{1\over q_T}\left\{ {P_h^\mu \over z_f} -q^\mu -\left(
1+{q_T^2+m_h^2/z_f^2 \over Q^2}\right)x_{bj} p^\mu\right\},\nonumber\\
Y^\mu &=& \epsilon^{\mu\nu\rho\sigma}Z_\nu X_\rho T_\sigma ,\nonumber\\
Z^\mu &=& -{q^\mu \over Q}. 
\eeq
These are the extension of four basis vectors
introduced for the massless case ($m_h=0$) in \cite{MOS92}
to the case of massive meson with $m_h$ in the final state\,\cite{KQ08}. 
These vectors become
$T^\mu=(1,0,0,0)$, $X^\mu=(0,1,0,0)$, $Y^\mu=(0,0,1,0)$, $Z^\mu=(0,0,0,1)$
in the hadron frame.  
In the present case, we find that $W^{\mu\nu}$ can be expanded in terms of
the following six independent tensors\,\cite{KT071}: 
\beq
&&{\cal V}_1^{\mu\nu}=X^\mu X^\nu + Y^\mu Y^\nu,\qquad
{\cal V}_2^{\mu\nu}=g^{\mu\nu} + Z^\mu Z^\nu,\nonumber\\
&&{\cal V}_3^{\mu\nu}=T^\mu X^\nu + X^\mu T^\nu,\qquad
{\cal V}_4^{\mu\nu}=X^\mu X^\nu - Y^\mu Y^\nu,\nonumber\\
&&{\cal V}_8^{\mu\nu}=T^\mu Y^\nu + Y^\mu T^\nu,\qquad
{\cal V}_9^{\mu\nu}=X^\mu Y^\nu + Y^\mu X^\nu, 
\eeq
where we have followed the notation in \cite{MOS92} 
but have not shown the explicit form of the
three tensors ${\cal V}_{5,6,7}$ among nine basis tensors 
because these three tensors are irrelevant for
the
expansion of our $W^{\mu\nu}$.
We also introduce
the inverse tensors $\widetilde{\cal V}_k^{\mu\nu}$ 
for the above ${\cal V}_k^{\mu\nu}$: 
\beq
&&\widetilde{{\cal V}}_1^{\mu\nu}={1\over 2}(2T^\mu T^\nu
+X^\mu X^\nu + Y^\mu Y^\nu),\qquad
\widetilde{{\cal V}}_2^{\mu\nu}=T^\mu T^\nu,\nonumber\\
&&\widetilde{{\cal V}}_3^{\mu\nu}=-{1\over 2}(T^\mu X^\nu + X^\mu T^\nu),\qquad
\widetilde{{\cal V}}_4^{\mu\nu}={1\over 2}(X^\mu X^\nu - Y^\mu Y^\nu),\nonumber\\
&&\widetilde{{\cal V}}_8^{\mu\nu}={-1\over 2}(T^\mu Y^\nu + Y^\mu T^\nu),\qquad
\widetilde{{\cal V}}_9^{\mu\nu}={1\over 2}(X^\mu Y^\nu + Y^\mu X^\nu).  
\eeq
Then, one obtains  
\beq
L_{\mu\nu}W^{\mu\nu}=  \sum_{k=1,\cdots,4,8,9} \left[ L_{\mu\nu}{\cal V}_k^{\mu\nu} \right]
\left[W_{\rho\sigma}\widetilde{\cal V}_k^{\rho\sigma}\right]
\equiv Q^2\sum_{k=1,\cdots,4,8,9} {\cal A}_k
\left[W_{\rho\sigma}\widetilde{\cal V}_k^{\rho\sigma}\right], 
\label{LW}
\eeq
where ${\cal A}_k \equiv L_{\mu\nu}{\cal V}_k^{\mu\nu}/Q^2$ is given by
\beq
{\cal A}_1&=&1+\cosh^2\psi,\nonumber\\
{\cal A}_2&=&-2,\nonumber\\
{\cal A}_3&=&-\cos\phi\sinh 2\psi,\nonumber\\
{\cal A}_4&=&\cos 2\phi\sinh^2\psi,\nonumber\\
{\cal A}_8&=&-\sin\phi\sinh 2\psi,\nonumber\\
{\cal A}_9&=&\sin 2\phi\sinh^2\psi. 
\label{Ak}
\eeq
By the expansion (\ref{LW}), 
the cross section for $ep^\uparrow\to eDX$
consists of the five structure functions associated with
${\cal A}_{1,2}$, ${\cal A}_3$, ${\cal A}_4$, ${\cal A}_8$
and ${\cal A}_9$, respectively, which have
different dependences
on the azimuthal angle $\phi$.

In closing this section, we summarize the prescription established for
calculating the twist-3 single-spin-dependent cross section that is 
generated by the three-gluon correlation
functions of the nucleon:
The corresponding differential cross section is given by (\ref{diffsigma}) 
with the expansion in the right-hand side of (\ref{LW}), in which
$W_{\rho \sigma}$ is given as (\ref{Wmunu}) 
using our factorization formula (\ref{wfinal}) for $w_{\mu \nu}(p,q,p_c)$.


\section{Result for the twist-3 cross section for $ep^\uparrow\to eDX$}

Using the formalism presented above,
we now obtain
the leading-order QCD formula for the single-spin-dependent cross section 
in the SIDIS, $ep^\uparrow\to eDX$, 
generated from the twist-3 three-gluon correlation functions $O(x_1,x_2)$ and $N(x_1,x_2)$
of (\ref{3gluonO}) and (\ref{3gluonN}),
as
\beq
&&
\hspace{-0.3cm}\frac{d^5\Delta\sigma
}{dx_{bj}dQ^2dz_fdq_T^2d\phi}\nonumber\\
&&
=\frac{\alpha_{em}^2\alpha_se_c^2 M_N}{8\pi
 x_{bj}^2S_{ep}^2Q^2}\left(\frac{-\pi}{2}\right) 
\sum_{k=1,\cdots, 4,8,9}
{\cal
 A}_k{\cal S}_k
\int_{x_{\rm min}}^1\frac{dx}{x}\int_{z_{\rm min}}^1\frac{dz}{z} \delta\left(
\frac{q_T^2}{Q^2}-\left(1-\frac{1}{\hat{x}}\right)\left(1-\frac{1}{\hat{z}}\right)
+\frac{m_c^2}{\hat{z}^2Q^2}\right)\nonumber\\
&&\qquad
\times \sum_{a=c,\bar{c}}
D_a(z) \left[\delta_a\left\{
\left(\frac{d}{dx}O(x,x)-\frac{2O(x,x)}{x}\right)\Delta\hat{\sigma}^{1}_k
+\left(\frac{d}{dx}O(x,0)-\frac{2O(x,0)}{x}\right)\Delta\hat{\sigma}_k^2
\right.\right.\nonumber\\
&&\qquad\qquad\qquad\qquad\left.\left.
+\frac{O(x,x)}{x}\Delta\hat{\sigma}^{3}_k
+\frac{O(x,0)}{x}\Delta\hat{\sigma}^{4}_k
\right\}\right. \nonumber\\
&&\qquad\qquad\left.
+\left\{
\left(\frac{d}{dx}N(x,x)-\frac{2N(x,x)}{x}\right)\Delta\hat{\sigma}^{1}_k
-\left(\frac{d}{dx}N(x,0)-\frac{2N(x,0)}{x}\right)\Delta\hat{\sigma}^{2}_k\right.\right.\nonumber\\
&&\left.\left.\qquad\qquad\qquad\qquad+\frac{N(x,x)}{x}\Delta\hat{\sigma}^{3}_k
-\frac{N(x,0)}{x}\Delta\hat{\sigma}^{4}_k
\right\}
\right],
\label{3gluonresult}
\eeq
where the subscript $k$ runs over $1,2,3,4,8,9$ with
${\cal A}_k$ defined in (\ref{Ak}) and ${\cal S}_k$ defined as
${\cal S}_k=\sin\Phi_S$ for $k=1,2,3,4$
and ${\cal S}_k=\cos\Phi_S$ for $k=8,9$.
The quark-flavor index $a$ can, in principle, be $c$ and $\bar{c}$, with $\delta_c=1$ and  
$\delta_{\bar{c}}=-1$, so that the cross section for 
the $\bar{D}$-meson production $ep^\uparrow\to e\bar{D}X$
can be obtained by 
a simple replacement of the fragmentation function
to that for the $\bar{D}$ meson, $D_a (z) \rightarrow \bar{D}_a (z)$. 
$\alpha_s=g^2/(4\pi)$ is the strong coupling constant, and
$e_c=2/3$ represents the electric charge of the $c$-quark.  
The lower limits of the integrals are given by
\begin{equation}
z_{\rm min}=z_f\frac{\left( 1 - x_{bj} \right)  Q^2}{2 x_{bj}m_c^2}  \left( 1
         - \sqrt{1- \frac{4x_{bj}m_c^2}{\left( 1 - x_{bj} \right) Q^2}
             \left[ 1
                + \frac{x_{bj} q_T^2}{\left( 1 - x_{bj} \right)  Q^2} \right]}\right) , 
\label{zmin}
\end{equation}
and
\begin{equation}
x_{\rm min} = \left\{ 
\begin{array}{ll}
x_{bj}\left[1+\frac{z_f^2q_T^2+m_c^2}{z_f(1-z_f)Q^2}\right]
& \quad \mbox{for}\;\;\; z_f\left(1+\sqrt{1+\frac{q_T^2}{m_c^2}}\right) > 1,\\ 
\\
x_{bj} \left[1+\frac{2m_c^2}{Q^2}\left(1+\sqrt{1+\frac{q_T^2}	 
{m_c^2}}\right)\right]
& \quad \mbox{for}\;\;\; z_f\left(1+\sqrt{1+\frac{q_T^2}{m_c^2}}\right) \leq 1.\\
\end{array} \right. 
\label{xmin}
\end{equation}
Partonic hard cross sections $\Delta\hat{\sigma}_k^i$ ($i=1,\cdots,4$)
are obtained as follows:
\beq
\left\{
\begin{array}{lll}
 \Delta\hat{\sigma}^{1}_1&=&{8q_T\hat{x} \over Q^6(1-\hat{z})^3\hat{z}^2}
\left\{Q^4\hat{z}
(1-\hat{z})(1-2\hat{z}+2\hat{z}^2-2\hat{x}+2\hat{x}^2+12\hat
{x}\hat{z}(1-\hat{x})(1-\hat{z})) \right.\nonumber\\[10pt]
&&\left.+2m_c^2Q^2\hat{x}(2\hat{z}(1-\hat{z})+\hat{x}(1-8\hat{z}+8\hat{z}^2))-4m_c^4\hat{x}^2\right\},
\\[10pt]
 \Delta\hat{\sigma}^{1}_2&=&\frac{64q_T\hat{x}^2}{Q^4(1-\hat{z})^2\hat{z}}
\left\{Q^2\hat{z}(1-\hat{x})(1-\hat{z})-m_c^2\hat{x}\right\}, \\[10pt]
 \Delta\hat{\sigma}^{1}_3&=&\frac{16\hat{x}}{Q^5(1-\hat{z})^3\hat{z}^3}(1-2\hat{z})
\left\{Q^2\hat{z}(1-\hat{x})(1-\hat{z})-m_c^2\hat{x}\right\}
\left\{Q^2\hat{z}(1-2\hat{x})(1-\hat{z})-2m_c^2\hat{x}\right\}, \\[10pt]
 \Delta\hat{\sigma}^{1}_4&=&\frac{32q_T\hat{x}^2}{Q^6(1-\hat{z})^3\hat{z}^2}
\left\{Q^2\hat{z}(1-\hat{x})(1-\hat{z})-m_c^2\hat{x}\right\}\left\{Q^2\hat{z}(1-\hat{z})+m_c^2\right\} ,
\\[10pt]
 \Delta\hat{\sigma}^{1}_8&=&\Delta\hat{\sigma}^{1}_9=0 ,
\end{array}
\right.\\
\label{sigma1}
\eeq
\begin{eqnarray}
\left\{
\begin{array}{lll}
\Delta\hat{\sigma}^{2}_1&=&\frac{8q_T\hat{x}}{Q^6(1-\hat{z})^3\hat{z}^2}\{Q^4\hat{z}(1-\hat{z})(1-2\hat{z}+2\hat{z}^2-4\hat{x}+4\hat{x}^2+24\hat
{x}\hat{z}(1-\hat{x})(1-\hat{z})) \\[10pt]
&&+4m_c^2Q^2\hat{x}(2\hat{z}(1-\hat{z})+\hat{x}(1-8\hat{z}+8\hat{z}^2))-8m_c^4\hat{x}^2\}, \\[10pt]
 \Delta\hat{\sigma}^{2}_2&=&2\Delta\hat{\sigma}^{1}_2, \\[10pt]
 \Delta\hat{\sigma}^{2}_3&=&2\Delta\hat{\sigma}^{1}_3 ,\nonumber\\[10pt]
 \Delta\hat{\sigma}^{2}_4&=&-\frac{16q_T\hat{x}}{Q^6(1-\hat{z})^3\hat{z}^2}(Q^2\hat{z}(1-2\hat{x})(1-\hat{z})-2m_c^2\hat{x})^2, \\[10pt]
 \Delta\hat{\sigma}^{2}_8&=&
\frac{16\hat{x}}{Q^3(1-\hat{z})^2\hat{z}^2}(1-2\hat{z})(Q^2\hat{z}(1-\hat{x})(1-\hat{z})-m_c^2\hat{x}) ,
\\[10pt]
 \Delta\hat{\sigma}^{2}_9&=&
-\frac{16q_T\hat{x}}{Q^4(1-\hat{z})^2\hat{z}}(Q^2\hat{z}(1-2\hat{x})(1-\hat{z})-2m_c^2\hat{x}),
\end{array}
\right.\\
\label{sigma2}
\end{eqnarray} 
\begin{eqnarray}
\left\{
\begin{array}{lll}
 \Delta\hat{\sigma}^{3}_1&=&\frac{16q_T\hat{x}^2}{Q^6(1-\hat{z})^3\hat{z}^2}(Q^2\hat{z}(1-2\hat{x})(1-\hat{z})-2m_c^2\hat{x})(Q^2(1-6\hat{z}+6\hat{z}^2)-2m_c^2) ,\\[10pt]
 \Delta\hat{\sigma}^{3}_2&=&-\frac{64q_T\hat{x}^2}{Q^4(1-\hat{z})^2\hat{z}}(Q^2\hat{z}(1-2\hat{x})(1-\hat{z})-2m_c^2\hat{x}), \\[10pt]
 \Delta\hat{\sigma}^{3}_3&=&-\frac{8\hat{x}}{Q^5(1-\hat{z})^3\hat{z}^3}(1-2\hat{z})
\left\{Q^4\hat{z}^2(1-\hat{z})^2(1-8\hat{x}+8\hat{x}^2)\right.
\nonumber\\[10pt]
&& \qquad\qquad\qquad
\left.  -8m_c^2Q^2\hat{x}\hat{z}(1-2\hat{x})(1-\hat{z})+8m_c^4\hat{x}^2\right\}, \\[10pt]
 \Delta\hat{\sigma}^{3}_4&=&-\frac{32q_T\hat{x}^2}{Q^6(1-\hat{z})^3\hat{z}^2}
(Q^2\hat{z}(1-\hat{z})+m_c^2)(Q^2\hat{z}(1-2\hat{x})(1-\hat{z})-2m_c^2\hat{x}),\\[10pt]
 \Delta\hat{\sigma}^{3}_8&=&
-\frac{8\hat{x}}{Q^3(1-\hat{z})^2\hat{z}^2}(1-2\hat{z})(Q^2\hat{z}(1-2\hat{x})(1-\hat{z})-2m_c^2\hat{x}),\\[10pt]
 \Delta\hat{\sigma}^{3}_9&=&
-\frac{32q_T\hat{x}^2}{Q^4(1-\hat{z})^2\hat{z}}(Q^2\hat{z}(1-\hat{z})+m_c^2) ,
\end{array}
\right.\\
\end{eqnarray} 
\begin{eqnarray}
\left\{
\begin{array}{lll}
 \Delta\hat{\sigma}^{4}_1&=&\frac{16q_T\hat{x}^2}{Q^6(1-\hat{z})^3\hat{z}^2}(Q^2\hat{z}(1-4\hat{x})(1-\hat{z})-4m_c^2\hat{x})(Q^2(1-6\hat{z}+6\hat{z}^2)-2m_c^2),\\[10pt]
 \Delta\hat{\sigma}^{4}_2&=&-\frac{64q_T\hat{x}^2}{Q^4(1-\hat{z})^2\hat{z}}(Q^2\hat{z}(1-4\hat{x})(1-\hat{z})-4m_c^2\hat{x}) ,\\[10pt]
 \Delta\hat{\sigma}^{4}_3&=&\frac{8\hat{x}}{Q^5(1-\hat{z})^3\hat{z}^3}(1-2\hat{z})
\left\{Q^4\hat{z}^2(1-\hat{z})^2(1+12\hat{x}-16\hat{x}^2)\right.\nonumber\\[10pt]
&&\left.\qquad\qquad\qquad\qquad +4m_c^2Q^2\hat
{x}\hat{z}(3-8\hat{x})(1-\hat{z})-16m_c^4\hat{x}^2\right\} ,\\[10pt]
 \Delta\hat{\sigma}^{4}_4&=&-
\frac{32q_T\hat{x}}{Q^6(1-\hat{z})^3\hat{z}^2}(Q^4\hat{z}^2(1-\hat{z})^2(1+\hat{x}-4\hat{x}^2)+m_c^2Q^2\hat
{x}\hat{z}(1-8\hat{x})(1-\hat{z})-4m_c^4\hat{x}^2),\\[10pt]
 \Delta\hat{\sigma}^{4}_8&=&\frac{8\hat{x}}{Q^3(1-\hat{z})^2\hat{z}^2}(1-2\hat{z})(Q^2\hat{z}(1+2\hat{x})(1-\hat{z})+2m_c^2\hat{x}),\\[10pt]
 \Delta\hat{\sigma}^{4}_9&=&-\frac{32q_T\hat{x}}{Q^4(1-\hat{z})^2\hat{z}}(Q^2\hat{z}(1-\hat{z})(1+\hat{x})+m_c^2\hat{x}) ,
\end{array}
\right.\\
\label{sigma4}
\end{eqnarray} 
where
\beq
\xhat={x_{bj}\over x},\qquad \zhat={z_f\over z}. 
\eeq

The single-spin-dependent cross section (\ref{3gluonresult}) 
can be decomposed into the five structure functions,
based on the different dependences on the azimuthal angles $\Phi_S$ and $\phi$ 
through the above-mentioned explicit forms of ${\cal A}_k$ and ${\cal S}_k$, as
\beq
\frac{d^5\Delta \sigma}{dx_{bj} dQ^2 dz_f dq_T^2 d\phi}
&=& \sin\Phi_S
\left( {\cal F}_1
+{\cal F}_2
\,{\cos\phi}
+{\cal F}_3
\,{\cos2\phi}\right)\nonumber\\
& &+\cos\Phi_S
\left({\cal F}_4
\,{\sin\phi}
+{\cal F}_5
\,{\sin2\phi}\right). 
\label{azimuth1}
\eeq
The five independent azimuthal structures of this type have been observed also in
the twist-3 single-spin-dependent cross section 
for $ep^\uparrow\to e\pi X$, generated from the quark-gluon correlation functions,
as presented in \cite{KT071,Koike:2009nc,Koike:2009yb}.  
Introducing the azimuthal angles $\phi_h$ and $\phi_S$ of the hadron plane 
and 
the nucleon's spin vector $\vec{S}$, respectively, as measured from the {\it lepton plane},
they are connected to the above $\Phi_S$ and $\phi$ as 
$\Phi_S=\phi_h-\phi_S$, $\phi=\phi_h$.
One can recast (\ref{azimuth1}) into the superposition of five sine 
modulations
with these new azimuthal angles as
\beq
\frac{d^5\Delta \sigma}{dx_{bj} dQ^2 dz_f dq_T^2 d\phi_h}
&=&\sin(\phi_h-\phi_S)\,F^{\sin(\phi_h-\phi_S)}
+\sin(2\phi_h-\phi_S)\,F^{\sin(2\phi_h-\phi_S)}
+\sin\phi_S\,F^{\sin\phi_S}\nonumber\\
&&+\sin(3\phi_h-\phi_S)\,F^{\sin(3\phi_h-\phi_S)}
+\sin(\phi_h+\phi_S)\,F^{\sin(\phi_h+\phi_S)},
\label{azimuth2}
\eeq
with the relations:
\beq
&&F^{\sin(\phi_h-\phi_S)}= {\cal F}_1
,\quad
F^{\sin(2\phi_h-\phi_S)}= \frac{{\cal F}_2 +{\cal F}_4}{2},
\quad
F^{\sin\phi_S}= \frac{-{\cal F}_2 +{\cal F}_4}{2},
\nonumber\\
&&F^{\sin(3\phi_h-\phi_S)}= \frac{{\cal F}_3 +{\cal F}_5}{2},
\quad
F^{\sin(\phi_h+\phi_S)}= \frac{-{\cal F}_3 +{\cal F}_5}{2}.
\label{Fsigma}
\eeq
The azimuthal-angle dependence of the single-spin-dependent cross section
derived in the TMD approach\,\cite{TMDsidis} was presented in a form similar to
(\ref{azimuth2}),
so that the decomposition (\ref{azimuth2}) into five structure functions 
is 
convenient to make connection with the TMD approach in the
small-$q_T$ region.

For completeness, we consider the 
unpolarized cross section for the SIDIS, $ep\to eDX$, and
list the corresponding twist-2 contribution at the leading order in QCD perturbation theory,
which gives an extension of the study in \cite{Mendez78,MOS92,KN03} to
the case with massive-hadron production in the final state.
The result is
\begin{eqnarray}
&&\hspace{-0.5cm}\frac{d^5\sigma^{\rm unpol}}{dx_{bj}dQ^2dz_fdq_T^2d\phi}
=\frac{\alpha_{em}^2\alpha_se_c^2}{8\pi
 x_{bj}^2S_{ep}^2Q^2}\frac{1}{4}
\sum_{k=1}^4{\cal
 A}_k\int_{x_{\rm min}}^1\frac{dx}{x}\int_{z_{\rm min}}^1 \frac{dz}{z}\sum_{a=c,\bar{c}}D_a(z)G(x)\hat{\sigma}^{U}_k\nonumber\\
&&\qquad\qquad\qquad
\qquad\qquad\qquad\times\delta\left(\frac{q_T^2}{Q^2}-
\left(1-\frac{1}{\hat{x}}\right)\left(1-\frac{1}{\hat{z}}\right)
+\frac{m_c^2}{\hat{z}^2Q^2}\right),
\label{unpolresult}
\end{eqnarray}
with (\ref{Ak}), (\ref{zmin}), and (\ref{xmin}).
This is generated from
the unpolarized gluon-density distribution
$G(x)$ for the nucleon, 
\beq
&&\hspace{-0.8cm}G(x)
=-\frac{g_{\beta \alpha}}{x}\int{d\lambda\over 2\pi}e^{i\lambda x}
\la p|F_a^{\beta n}(0)F_a^{\alpha n}(\lambda n)
|p\ra, 
\eeq
and
the partonic hard cross sections are obtained as, utilizing the formalism discussed
in section~3.3,
\begin{eqnarray}
\left\{
\begin{array}{lll}
 \hat{\sigma}^{U}_1&=&\frac{4}{Q^4(1-\hat{z})^2\hat{z}^2}\{Q^4\hat{z}
(1-\hat{z})(1-2\hat{z}+2\hat{z}^2-2\hat{x}+2\hat{x}^2+12\hat
{x}\hat{z}(1-\hat{x})(1-\hat{z})) \\[10pt]
&&+2m_c^2Q^2\hat{x}(2\hat{z}(1-\hat{z})+\hat{x}(1-8\hat{z}+8\hat{z}^2))-4m_c^4\hat{x}^2\},\\[10pt]
 \hat{\sigma}^{U}_2&=&
\frac{32\hat{x}}{Q^2(1-\hat{z})\hat{z}}(Q^2\hat{z}(1-\hat{x})(1-\hat{z})-m_c^2\hat{x}) ,\\[10pt]
 \hat{\sigma}^{U}_3&=&\frac{8q_T\hat{x}}{Q^3(1-\hat{z})^2\hat{z}}(1-2\hat{z})(Q^2\hat{z}(1-2\hat{x})(1-\hat{z})-2m_c^2\hat{x}), \\[10pt]
 \hat{\sigma}^{U}_4&=&\frac{16\hat{x}}{Q^4(1-\hat{z})^2\hat{z}^2}(Q^2\hat{z}(1-\hat{x})(1-\hat{z})-m_c^2\hat{x})(Q^2\hat{z}(1-\hat{z})+m_c^2) .
\end{array}
\right.
\label{sigmau}
\end{eqnarray}
We note that these partonic hard cross sections coincide with the corresponding
results presented in \cite{KQ08}, except for $\hat{\sigma}^U_1$.
Comparing (\ref{sigmau}) for $\hat{\sigma}^U_k$ with
the above result (\ref{sigma1}) for $\Delta\hat{\sigma}_k^1$, 
which represents the partonic hard cross sections associated with 
the derivatives, $dO(x,x)/dx$ and $dN(x,x)/dx$, of the three-gluon correlation functions 
in (\ref{3gluonresult}),
one finds the following relations between
the partonic hard cross sections at
the twist-3 level and those at the twist-2 level:
\beq
\Delta\hat{\sigma}_k^1={2q_T\xhat\over Q^2(1-\zhat)} \hat{\sigma}^U_k.
\label{unlike}
\eeq
The other partonic cross sections at the twist-3 level,
$\Delta\hat{\sigma}_k^i$ ($i=2,3,4$) of (\ref{sigma2})-(\ref{sigma4}),
are also related to the partonic cross sections (\ref{sigmau})
at the twist-2 level, although, unlike (\ref{unlike}), 
the corresponding relations cannot be manifested by direct comparison between the formulae
of (\ref{sigma2})-(\ref{sigma4}) and those of (\ref{sigmau}).
These remarkable relations, as well as a single ``master formula'' behind them,
will be presented elsewhere\,\cite{KTY10}.
We mention that the similar master formula,
which allows us to relate the 
$3\to 2$ partonic subprocess relevant for
the twist-3 level to the $2\to 2$ subprocess for the twist-2 level, 
was derived \cite{KT071} for 
the case of the SGP contributions
associated with the twist-3 quark-gluon correlation functions.

\section{Summary}

In this paper, we have investigated the single spin asymmetry
for the $D$-meson production in SIDIS, 
generated from 
the twist-3 three-gluon correlation functions for the nucleon.
We first showed, correcting the previous study in \cite{Ji92}, 
that there are only two independent three-gluon correlation functions 
of twist-3, 
$O(x_1,x_2)$ and $N(x_1,x_2)$, 
which correspond to two possible ways 
to construct color-singlet combination
composed of three active gluons.
Then, we have formulated the method for calculating the
twist-3 single-spin-dependent cross section generated from the three-gluon correlations.
Our formulation is based on a systematic 
analysis of the relevant diagrams in the Feynman gauge and 
gives all the contribution to the cross section at the twist-3 level in the leading order in perturbative QCD, 
guaranteeing the gauge invariance of the result.
As in the twist-3 mechanism for SSA
generated from the quark-gluon correlation functions for the nucleon, the cross section 
in the present case
occurs as the pole contribution of an internal propagator 
in the partonic hard-scattering subprocess, and the corresponding contribution leads to
the cross section expressed in terms of the
four types of functions of the relevant momentum fraction $x$:
$O(x,x)$, $O(x,0)$, $N(x,x)$ and $N(x,0)$.  
We find that all these four-types of functions and their derivatives with respect to $x$
contribute to the final form of the cross section.
These features discussed for the SSA in $ep^\uparrow\to eDX$ also apply to the
case of $A_N$ in $p^\uparrow p\to hX$ 
($h=\pi,\ K,\ D$, etc.)\,\cite{KY10}.

These new results have also revealed that
the previous studies \cite{KQ08,KQVY08}
missed important contributions.  
In particular, 
the factorization formula for the cross section used
in \cite{KQ08,KQVY08} was written down in an ad-hoc way, and,
compared to our factorization formula derived 
in this paper, 
involved additional projection 
for the Lorentz structure
onto a particular component
of the hard-scattering part as well as of the three-gluon correlation
functions,
which would have lead to the result incompatible with symmetry requirements in QCD.

\section*{Acknowledgments}
We thank Andreas Metz for bringing our attention to Ref.~\cite{BJLO}. 
The work of S. Y. is supported by the Grand-in-Aid for Scientific Research
(No. 22.6032) from the Japan Society of Promotion of Science.

\appendix

\section{Symmetry constraints 
on the gluon correlation functions}

In this Appendix, 
we consider the correlation functions of the gluon fields, $A_a^\mu(\xi)$,
in the nucleon.
Such correlation functions of the two- or three-gluon fields,
with non-lightlike separations between those fields,
arise in the intermediate step of the analysis of the relevant Feynman diagrams
for the SIDIS, $ep^\uparrow\to eDX$, to the twist-3 accuracy, as discussed in section~3.2.
We discuss the symmetry properties of those
correlation functions
and their implication 
on SSA: We show that Fig.~1(a) does not contribute to SSA, and that
Fig.~2(b) can give rise to SSA.

Similarly to (\ref{wtensor}), 
the contribution from Fig.~1(a) to $w_{\mu\nu}(p,q,p_c)$ in
(\ref{Wmunu}) can be written as,
\beq
\int{d^4 k\over (2\pi)^4}
S_{\mu\nu}^{(2)}(k)M_{(2)}^{\mu\nu}(k,p,S), 
\label{ms2}
\eeq
where the Lorentz indices for the virtual photon in the partonic hard-scattering part
$S_{\mu\nu}^{(2)}(k)$ are suppressed for simplicity, and 
the corresponding nucleon matrix element $M_{(2)}^{\mu\nu}(k,p,S)$ is defined as 
a correlation function of the type mentioned above:
\beq
M_{(2)}^{\mu\nu}(k,p,S)=\int\,d^4\xi e^{ik\cdot\xi}
\la pS| A_a^\nu(0)A_a^\mu(\xi)|pS\ra.  
\label{m2corr}
\eeq
Note that 
the color indices of the gluon fields in this formula are summed over,
since only the color-singlet combination is relevant as
a matrix element in the color-singlet hadron.    
Also, the corresponding color projection is taken 
for $S_{\mu\nu}^{(2)}(k)$ in (\ref{ms2}).
Now, 
invariance under the parity ($P$) and time-reversal ($T$) transformations implies
\beq
M_{(2)}^{\mu\nu}(k,p,S)^*=M_{(2)}^{\mu\nu}(k,p,-S),
\eeq
and, combined with the fact that the matrix element
$M_{(2)}^{\mu\nu}(k,p,S)$ depends on the spin vector $S^\mu$ {\it linearly},
we find that the spin-dependent part of
$M_{(2)}^{\mu\nu}(k,p,S)$ is a pure imaginary quantity.
For the process $ep^\uparrow\to eDX$, the leptonic tensor $L^{\mu\nu}$ of (\ref{dsigma}) is real.  
Accordingly, an imaginary contribution from the hard part $S_{\mu\nu}^{(2)}(k)$ is necessary
to give the real contribution to the spin-dependent cross section.  
However, this is impossible for the leading-order diagrams contributing to
the middle blob in Fig.~1(a).
Therefore, (\ref{ms2}) representing the contribution from Fig.~1(a)
does not give rise to SSA.

Next, using the similar logic as above, 
we consider the contribution of Fig.~1(b) to (\ref{wtensor}). 
By projecting $M_{abc}^{\mu\nu\lambda}(k_1,k_2)$ in (\ref{Mk1k20}) 
into the color-singlet components,
we define the two types of the contractions of the corresponding color indices 
as
\beq
M^{\mu\nu\lambda}_{\pm}(k_1,k_2,p,S)&&\equiv C^{bca}_\pm M_{abc}^{\mu\nu\lambda}(k_1,k_2)
\nonumber\\
&&=\int\, d\xi\int\, d\eta\, e^{ik_1\xi}e^{i(k_2-k_1)\eta}
\la pS | C^{bca}_\pm A_b^\nu(0)A_c^\lambda(\eta)A_a^\mu(\xi) |pS\ra ,
\label{Mk1k2}
\eeq
with $C_{\pm}^{bca}$ of (\ref{cpm}).
$PT$-invariance implies
\beq
M^{\mu\nu\lambda}_{\pm}(k_1,k_2,p,S)^*= M^{\mu\nu\lambda}_{\pm}(k_1,k_2,p, - S), 
\eeq
which shows that the spin-dependent parts of  
$M^{\mu\nu\lambda}_{\pm}(k_1,k_2,p,S)$ 
are pure imaginary quantities.  
Accordingly, 
only an imaginary contribution from the 
corresponding color-projected hard part 
$C^{bca}_\pm S_{\mu\nu\lambda}^{abc}(k_1,k_2,q,p_c)$
can give rise to SSA with (\ref{wtensor}) in $ep^\uparrow\to e DX$.  
For the diagrams shown in Fig.~1(b), 
the pole contribution of an internal propagator
in the hard part can give rise to such imaginary contribution.
The corresponding unpinched poles are allowed in Fig.~1(b),
owing to the presence of an extra gluon line connecting the hard and soft parts,
compared with Fig.~1(a).

At this point one might recall that the 
$k_\perp$-dependent gluon distribution function~\cite{MR01} (``gluon-Sivers function")
can give rise to SSA in the framework of the
TMD factorization and this fact may contradict with
the above statement that (\ref{m2corr}) does not contribute to SSA.  
However, the gluon Sivers function can become well-defined only after supplying 
an appropriate
gauge-link operator between the gluon-fields, and the gauge-link operator
actually resums the effect of the extra gluon-lines connecting the hard and soft
parts, i.e., the gluon-Sivers function represents some effect
of the contribution from Fig.~1(b). Therefore, there is no contradiction between the two facts.  
The gluon-Sivers function may rather represent the same effect
as the three-gluon correlation functions in the region of the intermediate transverse momentum,
similarly to the case of the quark-Sivers function and the twist-3 quark-gluon 
correlation functions as shown in \cite{JQVY06,JQVY06DIS,KVY08}.


\end{document}